\documentclass[12pt, preprint]{aastex}

\newcommand{\mo}{\ifmmode{\rm M_{\odot}}\else{M$_{\odot}$}\fi}
\newcommand{\lo}{\ifmmode{\rm L_{\odot}}\else{L$_{\odot}$}\fi}
\newcommand{\htw}{\ifmmode{\rm H_2}\else{H$_2$}\fi}
\newcommand{\htwo}{\ifmmode{\rm H_2O}\else{H$_2$O}\fi}
\newcommand{\mum}{\ifmmode{\mu m}\else{$\mu$m}\fi}
\newcommand{\gsim}{\ifmmode{{_{\sim}^{>}}}\else{$_{\sim}^{>}$}\fi}
\newcommand{\lsim}{\ifmmode{{_{\sim}^{<}}}\else{$_{\sim}^{<}$}\fi}
\newcommand{\ctht}{\ifmmode{\rm C_2H_2}\else{C$_2$H$_2$}\fi}
\newcommand{\iso}{{\em ISO}}
\newcommand{\iras}{{\em IRAS}}

\begin{document}

\title{Classification of 2.4--45.2 \mum\ Spectra from the \iso\ Short 
Wavelength Spectrometer\footnote{Based on observations with the {\em Infrared 
Space Observatory (ISO)}, an European Space Agency (ESA) project with 
instruments funded by ESA Member States (especially the Principle Investigator 
countries: France, Germany, the Netherlands, and the United Kingdom) and with 
the participation of the Institute of Space and Astronautical Science and the 
National Aeronautics and Space Administration (NASA).}
}

\author{Kathleen E. Kraemer,\altaffilmark{2,3} G. C. Sloan,\altaffilmark{4,5} 
Stephan D. Price,\altaffilmark{2} \and Helen J. Walker\altaffilmark{6}}

\altaffiltext{2}{Air Force Research Laboratory, Space Vehicles Directorate,
        29 Randolph Rd., Hanscom AFB, MA 01731; kathleen.kraemer@hanscom.af.mil, 
 steve.price@hanscom.af.mil}
\altaffiltext{3}{Institute for Astrophysical Research, Boston University,
        Boston, MA 02215}
\altaffiltext{4}{Institute for Scientific Research, Boston College, Chestnut 
Hill, MA 02467; sloan@ssa1.arc.nasa.gov}
\altaffiltext{5}{Infrared Spectrograph Science Center, Cornell 
University,    Ithaca, NY 14853-6801}

\altaffiltext{6}{Rutherford Appleton Laboratory, Chilton, Didcot, Oxon, OX11 
0QX, UK; H.J.Walker@rl.ac.uk}

%\slugcomment{Version 3:   \today} % ref/willner rewrite 1+co-a's

\begin{abstract}

The {\em Infrared Space Observatory} observed over 900 objects with the
Short Wavelength Spectrometer in full-grating-scan mode (2.4--45.2 \mum).
We have developed a comprehensive system of spectral classification
using these data.  Sources are assigned to groups based on
the overall shape of the spectral energy distribution (SED). The groups 
include  naked stars, dusty stars, warm dust shells, cool dust shells, 
very red sources, and sources with emission lines but no detected continuum.
These groups are further divided into subgroups based on spectral features 
that shape the SED such as silicate or carbon-rich dust emission,  
silicate absorption,
ice absorption, and fine-structure or recombination lines.
Caveats regarding the data and data reduction, and biases intrinsic to the
database, are discussed. We also examine how the subgroups relate to the
evolution of sources to and from the main sequence and how this 
classification scheme relates to previous systems.

\end{abstract}

\keywords{catalogs  --- stars: fundamental parameters ---
          infrared: ISM: continuum and lines ---
          infrared: stars --- ISM: general}

\section{Introduction}

Spectral classification organizes astronomical sources into groups with
similar properties based on the general or detailed morphology of their
spectral energy distributions (SEDs). Consequently, the classification 
criteria depend on the wavelength region and spectral resolution used. 
Both of these parameters must be uniform in order to create consistent 
criteria for arranging the sources in a database. The similarities and 
differences that result from applying a successful classification system 
to a sufficiently large sample of sources not only improve our knowledge 
about the sources but provide a basis for understanding the physical 
parameters of the objects.

The best example of how a classification system can lead to insight 
into the physical properties of the objects studied is provided by 
optical spectral classification \citep[cf.][]{hs86}.
  From the earliest systems based on 
general color \citep[e.g.][]{ruth63}, several competing systems emerged 
based on spectral line ratios \citep[e.g.][]{sec66,sec68,vog74,vw99,
pic90}.  Of these, the Harvard 
system used originally in the Draper Memorial Catalogue \citep{pic90}
 grew to predominate due to the large numbers of sources 
classified ($>10,000$), and served as the basis for the Henry Draper 
Catalogue \citep[beginning with][]{cp18}.

The MK spectral classification system evolved from the Harvard system 
\citep[e.g.][]{mor38,mkk43}. This 
two-dimensional system provided the clues necessary to disentangle the
different stages of the life cycle of a star and the relation of 
intrinsic parameters such as mass and metallicity to directly
observable properties.  MK spectral classification remains the 
single most powerful diagnostic tool available to astronomers when
applied to naked stars, i.e. stars not embedded in dust.

Unfortunately, the very early and very late stages of stellar evolution
rarely involve naked stars. The sources are deeply embedded within
interstellar dust clouds or circumstellar dust shells, either of which 
absorb the optical radiation and re-emit it in the infrared.  This dust can
absorb so much of the optical radiation from the star that traditional
classification based on the photospheric properties of the star in the 
optical is 
difficult, if not impossible.  Near-infrared observations can often penetrate 
the obscuring dust, permitting direct measurements of the stellar 
photosphere.  The spectral region between 1 and 9 \mum\ is rich in atomic
and molecular lines which trace temperature and luminosity.  For example, 
CO, SiO, and water vapor are sensitive indicators in oxygen-rich stars, even 
with low spectral resolution; the Phillips and Ballick-Ramsey C$_2$ bands 
as well as CN and CO serve for carbon stars.  However, the emission from 
the dust distorts the photospheric continuum and fills in the absorption 
features, making analysis difficult. Observations in the thermal infrared 
trace the emission from the dust itself.  The characteristic SED of the 
dust is distinctive enough to serve as the basis for classification 
\citep{lmp86,lm87,css89}.

The infrared spectra obtained by the Low-Resolution Spectrometer (LRS) 
on the {\em Infrared Astronomical Satellite} (\iras) are the best example of 
a nearly complete, self-consistent database that is ideal for spectral 
classification. These spectra cover wavelengths from 7.7 to 22.7~\mum\ 
at a spectral resolution of $\lambda/\Delta\lambda\sim$20--60. The
original LRS atlas contained  spectra from 5,425 sources \citep{lrs86}.
\cite{vk91} expanded the database to 6,267 and  \cite{kvb97} extracted 
almost 5000 additional spectra from the raw data to create a spectral 
database of 11,224 sources, making the LRS observations the largest infrared 
spectral database to date.  This database includes most of the 12 \mum\ 
objects in the sky brighter than 10 Jy at 12 \mum\ (magnitude +1),  and 
several infrared classification systems have been developed from it.

The initial LRS classification scheme \citep{lrs86,lrs88} sorted the 
original database of 5,425 sources into 10 groups, essentially  based 
on the dominant spectral feature in the 10 \mum\ region. 
These groups were subdivided further, 
usually by the strength of the dominant feature.  The AutoClass algorithm 
(also known as AI for artificial intelligence) used a Bayesian algorithm 
to sort the database into self-consistent classes with no a priori input 
about the nature of the spectra \citep{css89, gsv89}. \cite{kvb97}  
used one-letter 
codes to identify the character of each spectrum in the expanded database 
(11,224 sources).  These various classification systems have divided the LRS 
database into distinct sets of spectral classes. However, none of
these  systems has been applied to a substantial number of
spectra from instruments other than the LRS.

Other schemes focused on subsets of the LRS database.  For example,
\citet[][hereafter, LML]{lml88,lml90} classified evolved oxygen-rich stars 
based on their dust emission characteristics.  This system, as modified by 
\citet[][hereafter SP]{sp95,sp98}, has also been applied to ground-based 
spectral measurements \citep[e.g.][]{ce97, mgd98}.

Spectra  taken by the Short Wavelength Spectrometer (SWS) on the {\em
Infrared Space Observatory} \citep[\iso][]{kes96,deg96} are now publically
available. In this paper, we focus on the full-range, moderate-resolution 
spectra obtained in the SWS01 observing mode. These observations are over a
greater wavelength range than the LRS database (2.4--45.2 \mum\
compared to 7.7--22.7 \mum), and at a higher spectral resolution
($>$300--400 vs. 20--60).  The SWS01 spectral resolution is sufficient 
for detailed examination of band structure and atomic fine-structure
lines.  The extended wavelength range includes both the near-infrared 
spectral region, which is dominated by molecular bands from stellar 
photospheres, and the thermal infrared region, which is dominated by 
dust emission.  The LRS database is compromised by inadequate
wavelength coverage on the short-wavelength side of the strong
spectral features produced by silicate dust (10~\mum) and
silicon-carbide grains (11.5~\mum), making it difficult 
unambiguously define the stellar continuum.

\iso\ obtained observatory-style pointed observations whereas \iras\
obtained spectra as an adjunct to the main survey with the LRS as a 
secondary instrument.  Consequently, the SWS database only contains full-range 
spectra of $\sim910$ specifically targeted sources (1248 total spectra, 
including duplicates and off-positions).  To ensure that \iso\ obtained SWS
spectra of as wide a variety of sources as possible, the observing lists of 
the STARTYPE
proposals\footnote{The STARTYPE proposals received \iso\ project names
STARTYP1, STARTYP2, and ZZSTARTY.} targeted sources in categories which were
 under-represented in the infrared classification 
systems (\S \ref{sec.sample}).  The result is a robust database of 
infrared spectra which  is the basis for our infrared spectral classification
system.

We describe the sample of the observed sources and the structure and 
calibration of the spectral data in Section 2.  Section 3 details the 
criteria for the classification system, which we discuss in Section 4.  
The actual classifications are presented in Appendix A.

\section{Observations and Data Analysis \label{sec.obs}}

\subsection{The Sample \label{sec.sample}}

\subsubsection{Source Selection}
The SWS database contains observations obtained for a wide range of 
individual observing projects.  A series of observing proposals, 
referred to collectively as the STARTYPE proposals, was 
developed to supplement observations from other dedicated and open-time
experiments. The original observing lists  included at least one source from
each category defined by the MK spectral types, LRS classifications,
AutoClass classifications, and the spectral templates in the Galactic
Point Source Model \citep{wcv92}. While the MK classification system is 
familiar to most readers, the infrared classification systems may be 
less so. Therefore, we describe below the three infrared 
classification systems used to create the STARTYPE observation lists.

The LRS classifications presented in the LRS atlas \citep{lrs86}
used a two-digit scheme to describe a spectrum.  The scheme
subjectively divided the spectra into 10 groups (identified by the
first digit) based on spectral morphology and, in part, on ideas about
the underlying physics producing the spectra.  For blue sources 
(i.e. flux decreasing with wavelength),
 the first digit represents (1) featureless spectra, (2) silicate 
emission, (3) silicate absorption, or (4) carbon-rich dust emission 
features. Red sources were assigned a first digit of 5, 6, or 7 
(analogs of 1, 2, or 3).  Spectra dominated by emission lines were 
divided into (8) those with unidentified infrared (UIR) bands  and 
(9) those without UIR bands. Miscellaneous spectra were assigned an 
initial digit of 0.  The second digit was typically based on
the strength of the features identified by the first digit.  In
general, the spectral morphology is clearly different among the groups,
but some inconsistencies and misclassifications exist.

The AutoClass scheme  \citep{gsv89,css89} used artificial intelligence  to 
sort the LRS spectra into a series of self-consistent classes.
By separating features on the basis of both shape and strength, 
this method distinguished subtleties not addressed by the LRS atlas
characterizations, which separated features based on strength alone.
It also found weak features that had been previously undetected. 

Selecting sources only from classification systems based on the LRS database
excludes information about the wavelength regions not covered by the LRS,
that is, shortward of 7.7 \mum\ and longer than 22~\mum.  To address this 
shortcoming, the
STARTYPE experiments also selected sources based on the spectral templates in 
the Galactic Point Source Model \citep{wcv92}. Sources in this scheme are 
divided into classes based either on the MK spectral classifications or 
location in the $[12]-[25]$, $[25]-[60]$ color plane.  \cite{cww90}
grouped sources in the $[12]-[25]$, $[25]-[60]$ plane and created
prototypical spectral templates  used in the \cite{wcv92}
Galactic Point Source Model.

An initial list of 1316 sources brighter than 40 Jy at 12 \mum\ was compiled  
by randomly selecting $\sim10\%$ of the sources that were used to create the 
classification schemes described above. Because all three infrared schemes 
use the LRS spectra, roughly 25\% of the sources were randomly selected from 
the LRS atlas, then sorted by LRS class.  When a particular LRS sub-class was 
well-populated after the initial selection, sources  associated with objects 
in other catalogs were preferentially chosen due to the additional information 
available on them.  Because the 12 \mum\ flux criterion discriminated
against red and emission line objects, the flux limit was lowered to 5 Jy 
at 12 \mum\ to include 347 red objects (LRS classes $5n$ through $9n$) and to 
increase the number of sources in other under-populated LRS classes.  The 
resulting list was tabulated in terms of the number of sources in each LRS 
and AutoClass sub-classes.  If a sub-class had more than ten objects in the 
list, objects with higher quality LRS spectra were preferentially selected. 
Noisy counterparts of other, better defined, sub-classes and faint, 
unique classes (such as 01 or $\theta0$) were proportionately 
under-represented 
in the observing list. The final 10\% observing list was comprised of somewhat
less than 800 sources. 

Given the time constraints of the \iso\ mission, only a fraction,
approximately one tenth, of the 10\% list could be observed.
The initial STARTYPE
observing list sparsely, but uniformly, sampled the LRS and AutoClass
subclasses (Table \ref{tab.irascnts}) and populated the MK classes. 
Because  astrophysically interesting sources included in our list 
would likely be observed by other experimenters, we deferred the majority 
of our observations until we had surveyed the observing lists for the 
dedicated-time and open-time SWS01 spectra to determine which spectral
classes were under-represented.  As expected, the SWS01 observing 
lists of other experimenters sampled the red LRS classes (5$n$ 
through 9$n$) and the equivalent AI classes well. The other classes were not 
as well sampled. We then concentrated the STARTYPE observations on the 
types of objects not observed by other \iso\ investigators, such as 
those with featureless continua and carbon stars with 
small circumstellar excesses. Therefore, additional sources within LRS 
classes $1n,~2n,$ and $4n$ were included in the STARTYPE observing list 
to provide more spectral representatives, particularly for the important 
subclasses 29, 43 and 44.  Several LRS class $3n$ sources were also 
included to correct the slight under-representation across the entire 
class. In this sample, stars of MK luminosity class V are well 
represented from B to early G, IIIs from early G to late M, and there is 
a sprinkling of temperature classes for the Is and IIs.  Because M dwarfs 
are faint, we used the PHT spectrometer PHT-S to obtain spectra of 6 
sources in the M dwarf sequence (Price et al. in preparation).

The observing scheme worked well.  Of the $\sim910$ individual sources with 
SWS01 spectra, 275 were among the 800 sources in our 10\% list. This 
number is increased to 379 if we had chosen to populate our subclasses to 
the 10\% limit with the sources actually observed, although the coverage is 
not as uniform (see below). Table \ref{tab.irascnts} shows how the number of 
sources observed compares to the number proposed, and to the total number of 
objects in each LRS and AutoClass subclass.

\subsubsection{Selection Effects \label{sec.seleff}}

 Although the STARTYPE program aimed at producing a
uniform sample, other programs did not.  Objects in most programs
were chosen with a particular research objective in mind, to investigate a
particular phenomenon or  a specific source.  Sources
with unusual features, intrinsically more interesting than sources
which are more typical, were observed more often than they would
have been in a completely uniform sample.  For example, $\eta$
Carina, as a unique object, would likely not have been observed in a
randomly selected sample of the 1248 objects, but was observed twice with SWS.
However, the tendency to observe more unusual sources
makes it more likely that the grid of subgroups for classification
includes most of the possible types of infrared spectra.

Comparing the number of objects in each LRS and AI class with the number 
actually observed (Table \ref{tab.irascnts}) provides some insight into 
the bias of the SWS database.  Although 47\% of the STARTYPE 10\% list 
was observed, the coverage of the \iras\ classes (either LRS
or AI) was significantly less uniform than the STARTYPE selections.
For example, STARTYPE proposed to observe 34 of the 324 objects
(10\%) in LRS class 14, but the SWS database includes only 7 (2\%).
This apparently uninteresting group consists of nominally naked stars
with spectral indices of $\beta\sim2$ \citep{lrs88}; in reality, these
sources exhibit low-contrast dust emission (see LML and SP).   In
contrast, a group of truly naked stars with  high signal-to-noise 
defined by AutoClass $\delta0$ \citep{css89,gsv89} had 44 of 256 objects 
(16\%) observed instead of the 27 suggested by STARTYPE, primarily because
this group included the chosen calibration stars.  Roughly 15\% of the LRS 
and AutoClass
classes had significantly fewer sources observed than if the
STARTYPE sample had been followed, including several which in the end
had no members observed.  On the other hand, the SWS database
includes more than twice as many PNe and star forming regions,
source types which include the brightest objects in the infrared
sky, than does the LRS Atlas \citep{lrs86}, significantly expanding the
available database on these important object types.

\subsection{Data from the \iso\ Data Archive \label{sec.data}}

The SWS obtained  1248 SWS01 spectra of over 900 different
sources.  The SWS01 spectra\footnote{Hereafter, the set of 
1248 SWS01 spectra are referred to as ``the SWS database.''}
cover wavelengths from 2.4 to
45.2~\mum\ in 12 spectral segments (or bands).  The bands vary in
length from 0.2~\mum\ (Band 1A:  2.4--2.6~\mum) to over 16~\mum\
(Band 4:  29--45~\mum).  Each includes data from 12 individual
detectors taken in two scan directions (``up'' and ``down'' scans), giving 
a total of 24 discrete
spectra in each spectral segment.  Thus, to produce one full-scan
spectrum from the SWS, 288 individual spectra must be calibrated and
combined. 

The standard ``basic science'' format for SWS spectra from the 
\iso\ Data Archive (IDA) is the 
Auto-Analysis Result (AAR) produced by the Off-Line Processing (OLP) 
pipeline.  To classify the spectra, we typically used the browse product, 
which was created from OLP version 7.1.  The browse product collapses the 
individual spectral scans to one usable spectrum, which usually sufficed 
for classification.  For problematic spectra, we used a preliminary release 
of OLP version 10.0, combining the data into one spectrum using software 
written at the Air Force Research Laboratory. Sloan et al. (2002) will 
present further details of this method, as well as a spectral atlas of
the 1248 spectra.

Despite efforts to calibrate the flux of each spectral segment in the
standard pipeline, discontinuities often exist at the boundaries between 
each of the 12 bands \citep{skp01,ksp01, ship01}. For compact sources 
(smaller than the aperture), this problem most likely 
results from errors in satellite pointing \citep{ship01}.  Since the point 
spread function (PSF) is comparable to the angular size of the aperture, a 
slight offset from the center of the aperture will truncate the PSF.  

A formal solution to the discontinuities does not yet exist, but a 
work-around has produced satisfactory results.  Although the bands have 
sharply defined edges, adjacent bands include overlap regions of 
$\sim$0.15--2.0~\mum.  While only data from one band was considered to 
be ``in-band'' for a particular overlap region, with a well-calibrated 
relative spectral response function (RSRF), the ``out-of-band'' data can 
often be used to verify spectral features, the shape of the SED in that 
overlap region, and, most importantly, the flux level. To correct for the 
band-to-band discontinuities, the flux from a (usually) well-behaved spectral 
segment was chosen to be the fiducial segment and the other 
segments normalized to it, usually by a multiplicative factor. 
An additive factor was used for fainter sources where dark current
variations might dominate gain variations. The same band could not be used 
for all sources due to the lack of flux in that band for certain SEDs.
Band 1B (2.60--3.02 \mum) served as the fiducial segment for sources 
dominated by flux from the stellar photosphere. For red sources peaking 
beyond $\sim$15~\mum, Band 3C (16.5--19.5 \mum) was the fiducial segment.

The detectors of Bands 2 (4.08--12.0 $\mu$m, Si:Ga) and 4
(29.0--45.2~\mum\ Ge:Be) exhibit memory effects which can lead to 
differences in signal between the up and down scans 
\citep[cf.][]{skp01,ksp01}.  This problem manifests itself as a
variation in dark current during a scan, the magnitude of which depends 
on the recent flux history of the detector.
The SWS Interactive Analysis\footnote{The SWS Interactive Analysis
system is developed and maintained by the SWS consortium members
(Space Research Organization of the Netherlands, Max Planck
Institut f\"ur Extraterrestrische Physik, Katholiede
Universiteit Leuven, and the European Space Agency).} (IA)
routine {\it dynadark} was developed to model the dark current in Band 2.
The algorithm in this routine is based on the Fouks-Schubert formalism %
\citep{fs95,fou01,kfl01}, which accounts for non-linear responses in the
flux history of the detectors. In its current form, this routine behaves
erratically.  
It can substantially improves the Band 2 data, but it can also overcorrect 
the data, degrading the match between up and down segments in Band 2A or 2B
\citep{skp01,ksp01}.  The Pipeline 10 processing automatically includes 
the {\it dynadark} routine.

The memory effect in Band 4 affects the shape of the spectrum
longward of $\lambda\sim$38--40~\mum.  The degree to which a
spectrum is affected depends on the underlying SED of the source and
its brightness, as well as when during the mission and at what speed
the observation was made.  Changes in the calibration strategy
involving photometric checks and dark current measurements by
revolution 200 helped to some extent.   However, the underlying
problem with the memory effects remains unsolved.

Discontinuities in the 26--30~\mum\ region are caused by a
combination of changing aperture size (especially for extended
sources), pointing issues, a light leak in Band 3D, and the
poor behavior of Band 3E in many spectra.  The first two problems
require multiplicative corrections (to first order).  The light leak
appears at the long-wavelength end of Band 3D and results from radiation
from Band 3A leaking through the filter for Band 3D.  When it occurs,
it invalidates data in Band 3D beyond $\sim$27.3~\mum.  Band 3E
often contains very noisy data, especially at fainter flux levels.
While the boundary between Bands 3D and 3E is officially 27.5~\mum\
\citep{lee01}, the former cannot be used beyond 27.3~\mum\ due to
the light leak and the latter is invalid below 27.7~\mum.  As a
result, normalization of one band to the other requires extrapolation
of the data in the gap between them.  Fortunately, Band 4 provides
relatively reliable data at wavelengths down to 27.7~\mum\ in OLP
10.0, even though its official cut-off is 29.0~\mum.  This extension
allows normalization of Band 4 directly to Band 3D by extrapolation, 
bypassing the unreliable data in Band 3E.

The procedure used to reduce the band discontinuities assumes that
the flux levels for Band 1B or 3C are reliable.  Any errors in the
absolute flux level within those bands will be propagated to the
other bands through the normalization process.  Furthermore, if the
wrong reference band was chosen or an overlap region is unusually
noisy, incorrect normalization can degrade the data.  This
problem is especially acute for weak sources with peak fluxes less
than $\sim$25 Jy. The impact of these calibration issues on the 
classification effort is discussed in \S \ref{sec.caliss}. 

\subsection{Classification Method \label{sec.method}}

We created a list of all SWS01 observations, regardless of object type or 
quality flag, from the IDA, giving a total of 1248 
spectra.  The browse product spectrum for each of the 1248 observations 
was examined for quality.  If a spectrum had no apparent signal, it was set 
aside; this included most observations designated as off or reference 
positions by the observer.  (If a spectrum had a discernible signal, it 
remained in the sample regardless of designation, such as the observation
originally designated M17NOFF.)  Roughly 35 objects contained no signal 
because the observer entered incorrect coordinates.  The OLP software flagged
an additional 30 or so spectra as having instrumentation,
telemetry, pointing, or quality problems, but we classified them
anyway.

Two of the authors (KEK and GCS) classified the sources independently, 
without prior knowledge of the MK spectral type, LRS class, or 
AutoClass category.  The separate classifications were then compared and 
combined 
into a single scheme.  Sources for which the placement was uncertain or 
unclear were reprocessed and re-examined.  Typically this reprocessing 
resolved the ambiguity, although often the assigned classification included 
a ``:'' or ``::'' to indicate uncertainty (see below).

\section{The Classifications \label{sec.class}}

We established three levels of classification\footnote{Use of  existing
LRS classifications was considered. However, those classes were based on
a limited spectral range and focused too strongly on the strength of
a single feature. This scheme was rejected as  inadequate to
describe the full range of SWS spectra (see \S \ref{sec.diff}).}.  The Level 1
categories (hereafter ``groups'') are sorted based on the general 
morphology of the SED, which is determined primarily by the temperature 
of the strongest emitter (be it stellar or dust).  Level 2 classification 
places each spectrum into a self-consistent subgroup based on the presence 
of prominent spectral features, such as silicate dust emission or 
absorption, carbon-rich dust emission, or atomic fine-structure lines.
\footnote{Acceding to a request by the editor to name our classification 
scheme, we hereby christen it ``the KSPW system.''}

Level 3 classification will be the arrangement of spectra within a 
given subgroup into a sequence. This is a complex, interactive
project left for the future.  Some studies have reached Level 3
classification for certain subgroups already well defined by previous
spectral databases.  For example, \cite{sp95,sp98} have already
developed a sequence for oxygen-rich dust spectra produced by
optically thin circumstellar shells.

\subsection{Level 1 Classification \label{sec.l1}}

The Level 1 classes  primarily depend on the temperature of the
dominant emitter.  Five main categories emerged, ranging from the
hottest objects such as naked stars (1) to the coolest objects such as
protostellar cores (5).  Additional categories 
include spectra with emission lines but no detected continuum (6) and
spectra which either contain no classifiable flux or are flawed for
some other reason (7).

\begin{enumerate}

\item  Naked stars.  Photospheric emission with no apparent influence
from circumstellar dust dominates these spectra.  All sources have
optical identifications with known and well-classified stars.

\item  Stars with dust.  The SEDs are primarily photospheric at
shorter wavelengths, but they also show noticeable or significant
dust emission at longer wavelengths.  Most sources are red
supergiants or are associated with the asymptotic giant branch (AGB).

\item Warm, dusty objects. These sources are dominated by emission from warm 
dust.  Any photospheric contribution from an embedded star is either absent 
or significantly less than the peak emission.  The emission typically 
peaks between $\sim$5 and $\sim$20~\mum, which implies dust temperatures 
hotter than $\sim$150 K.  The majority of these spectra arise from deeply 
enshrouded AGB sources, transition objects, planetary nebulae (PNe), or 
other evolved sources.

\item  Cool, dusty objects.  These objects are dominated by cooler
dust emission, the peak of which occurs within the SWS spectral range but
longward of $\sim$20~\mum.  Most sources in this group are PNe, AGB 
stars, and transition objects, although many are young stellar
objects (YSOs).

\item  Very red objects.  These objects have rising spectra toward 
longer wavelengths through at least the end of Band 4.  Most sources 
are star-forming regions or PNe. 

\item  Continuum-free objects but with emission lines.  These sources
do not have enough continuum emission to allow an unambiguous
placement in another group.  Emission lines, typically atomic
fine-structure lines, dominate the spectra.  Objects in this group
include supernova remnants and novae.  Because these spectra are
often difficult to discern from the class 7 spectra, this group may
not contain all possible members observed by \iso.

\item  Flux-free and/or fatally flawed spectra.  This group includes 
objects with no detected flux or flux levels insufficient for 
classification.  In addition to intrinsically faint objects, this
group contains observations with incorrect coordinates, observations 
intentionally offset from sources (off-positions), and flagged 
observations.  Spectra with enough flux to allow classification
appear in the appropriate group whenever possible despite flags or
off-target coordinates.

\end{enumerate}

\subsection{Level 2 Classification \label{sec.l2}}

Level 2 classification separates the Level 1 groups into subgroups
based on the spectral features superimposed on the overall SED.
Each subgroup has a one-, two-, or three-letter designation which 
 succinctly indicates  the type of dust and most prominent
feature(s), as described below and summarized in Table
\ref{tab.abbrev}.  In addition to the letter designations,
one-character suffixes describe any unusual properties of the
spectrum (Table \ref{tab.suf}).

The initial letter of the designation indicates the overall
``family'' to which an object belongs.  The three most important
families are ``S'', ``C'', and ``P.''  ``S'' indicates oxygen-rich
dust material such as silicate or alumina grains, whereas ``C''
indicates carbon-rich material.  ``P'' indicates planetary nebulae
(PNe), which typically have spectra rich in emission lines.

The second and third letters, if used, indicates the presence of one
or more specific spectral features.  The letter combinations present
in Groups 2--5  are:

\noindent{\bf SE}---Silicate or oxygen-rich dust emission feature at
            $\sim$10--12~\mum, usually accompanied by a secondary
            emission feature $\sim$18--20~\mum.

\noindent{\bf SB}---Self-absorbed silicate emission feature at 
            10~\mum, usually showing emission peaks at 9 and 11~\mum.
            The secondary emission feature $\sim$18--20~\mum\ is common.

\noindent{\bf SA}---Silicate absorption feature at 10~\mum.  The
            18--20~\mum\ feature can be in emission or absorption.
            Features from crystalline silicate emission may also be
            present at longer wavelengths.

\noindent{\bf SC}---Crystalline silicate emission features, especially
            at $\sim$33, 40, and/or 43~\mum.  No significant silicate 
            features apparent $\sim$10~\mum.

\noindent{\bf SEC}---Crystalline silicate emission features, 
            especially at 11~\mum, usually at $\sim$19, 23, and
            33~\mum, and often at 40 and/or 43~\mum.  The presence of
            crystalline silicates has shifted the emission feature at
            10~\mum\ due to amorphous silicate grains $\sim$1~\mum\
            to the red.  The presence of other crystalline features
            distinguishes this feature from the self-absorbed
            silicate emission (SB) feature, which also peaks
            $\sim$11~\mum.

\noindent{\bf CE}---Carbon-rich dust emission dominated by the
            silicon carbide emission feature at $\sim$11.5~\mum.
            The shape and wavelength of this feature differs 
            substantially from the SB and SEC features at 11~\mum,
            and any uncertain cases can be resolved by the 
            presence of a narrow absorption feature at 13.7~\mum\
            \citep[due to \ctht; e.g][]{ato99, cyg99, vxk00}.

\noindent{\bf CR}---Carbon-rich dust emission showing a reddened
            continuum (due to a strong contribution from amorphous
            carbon), the SiC emission feature at $\sim$11.5~\mum, and
            another emission feature at $\sim$26--30~\mum.

\noindent{\bf CT}---Carbon-rich dust emission characterized by a red
            continuum and emission features at 8, 11.5, 21,
            26--30~\mum.  The ``T'' stands for the
            ``{\bf T}wenty-one''\mum\ emission feature, which is the
            primary discriminant between CR and CT.

\noindent{\bf CN}---Carbon-rich proto-planetary nebulae with
            11.5~\mum\ emission or the 13.7 \mum\ absorption features,
            and much redder SEDs as compared to the CRs.

\noindent{\bf C/SC}---Carbon-rich features in the blue half of the
            spectrum, combined with crystalline silicate emission
            features at 33, 40, and/or 43~\mum.

\noindent{\bf C/SE}---Carbon-rich features in the photospheric emission,
            combined with silicate or oxygen-rich dust emission 
            at $\sim$10--12~\mum. These are the silicate carbon stars 
            \citep[e.g.][]{lm86,le90}.

\noindent{\bf PN}---Prominent emission lines from atomic
            fine-structure transitions.

\noindent{\bf PU}---Similar to PN, but with strong UIR
            features as well (see below).

\noindent{\bf U}---Prominent emission features at 3.3, 6.2, 
            $\sim$7.7--7.9, 8.6, and 11.2~\mum\, commonly described as
            UIR features.  They most
            likely arise from polycyclic aromatic hydrocarbons
            (PAHs), although this identification remains
            controversial.  Unless specified otherwise, there are no
            other strong spectral features.  Sources with
            low-contrast UIR emission difficult to detect when
            examining full-scan spectra may not be classified as
            ``U''.  In other words, many sources with fainter UIR
            features are classified in other groups.

\noindent{\bf U/SC}---A combination of UIR emission features in the
            blue half of the spectrum and crystalline silicate
            emission features at 33, 40, and/or 43~\mum.

\noindent{\bf E}---No discernible spectral structure, except for the
            presence of atomic emission lines.

\noindent{\bf F}---Featureless spectrum (within the signal/noise
            ratio).

\noindent{\bf W}---The continuum emission peaks $\sim$6--12 \mum,  
            usually with apparent silicate absorption at 10~\mum.
            The ``W'' stands for Wolf-Rayet, since these spectra are
            always produced by Wolf-Rayet stars or R Corona Borealis
            variables.
 
\noindent{\bf M}---Miscellaneous spectra:  most of these objects have
            distinct features but could not be placed in any of the
            other existing categories, even with a ``p'' suffix.
            Objects that clearly belong in the parent group but are
            too noisy to classify further into a subgroup  also
            appear here.

Because some spectral characteristics occur across a broad range of 
temperature, the same Level 2 subgroup description can appear in
different Level 1 groups.  Table \ref{tab.cross1-2b} summarizes the
occurrence of each Level 2 subgroup within the Level 1 groups.

\subsection{Group Descriptions and Sample Spectra \label{sec.groups}}

Each group of Level 1 spectra separates into several subgroups,
often including a subgroup for peculiar or noisy spectra which
defied attempts to unabiguously place them elsewhere.  The
figures illustrate sample spectra for each subgroup.  Spectral
classifications and source types are taken from the literature
or from SIMBAD.

\subsubsection{Group 1---Naked Stars \label{sec.g1}}

\begin{figure}
\plotone{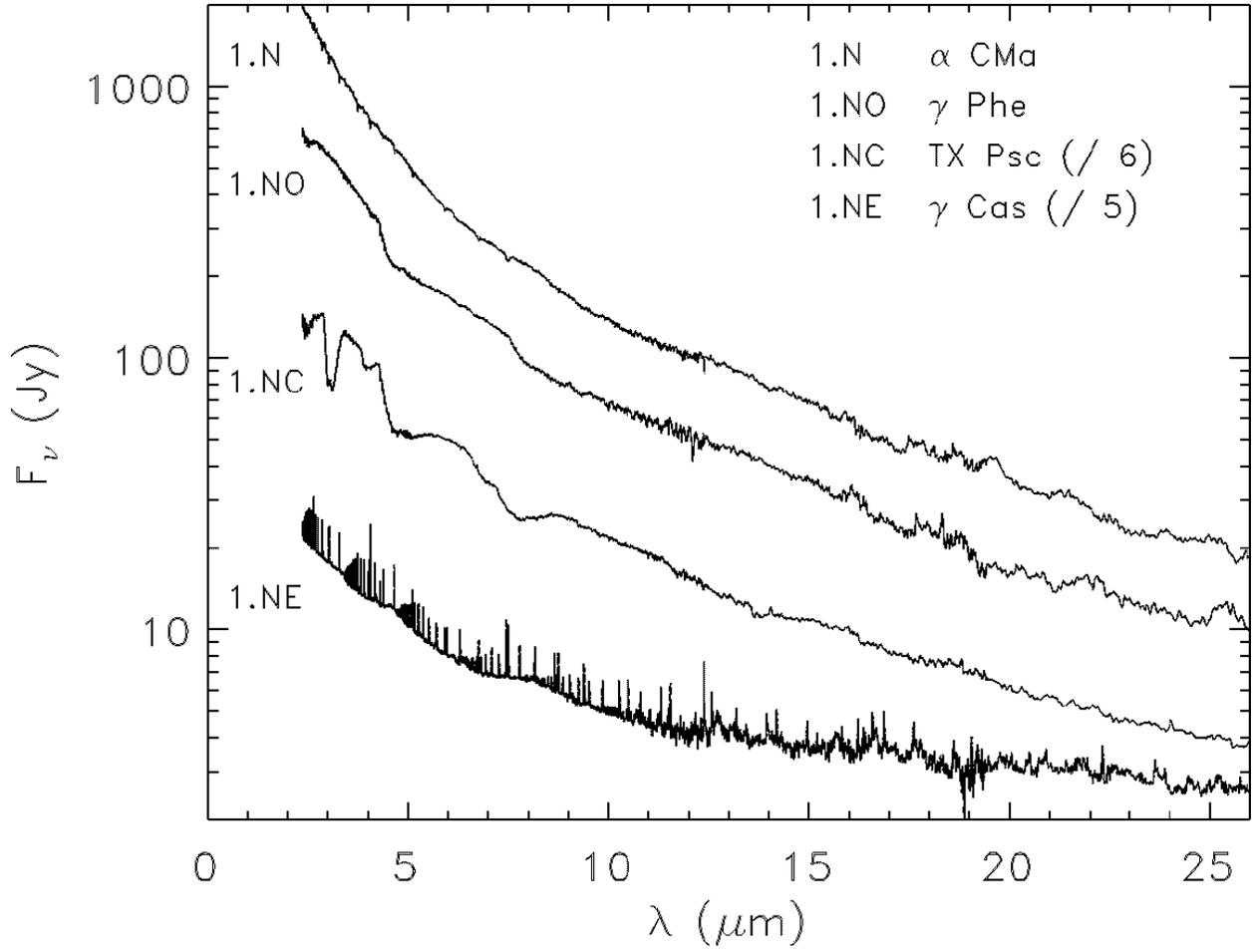}
\caption{Typical spectra from Group 1.  Numbers in parentheses after the
object name indicate the scaling factor used to make the plots.
}
\label{fig.1}
\end{figure}

The naked stars fall into several easily distinguished subgroups
based primarily on the presence or absence of molecular absorption
bands.  These include ordinary stars (1.N), oxygen-rich stars (1.NO),
carbon-rich stars (1.NC), and emission line stars (1.NE).  An
additional subgroup (1.NM) includes sources whose SEDs are dominated by 
photospheric emission, but are
too noisy or otherwise too peculiar to place with confidence in one
of the main subgroups.  Figure \ref{fig.1} presents examples of each
subgroup.

\noindent{\bf 1.N}  The 1.N stars include  the main sequence
stars with no molecular bands in their spectrum.  A combination of a
simple Engelke function \citep{e92} and narrow atomic absorption
features (primarily hydrogen recombination lines) accurately
describes the spectrum. MK classifications of stars in this subgroup
range from O9V ($\zeta$ Oph) to K0Iab ($\alpha$ UMi).

\noindent{\bf 1.NO}  The 1.NO stars show broad absorption features
in their spectra from the CO overtone (maximum absorption
$\sim$2.5~\mum), a blend of the SiO overtone (4.2~\mum) and the CO
fundamental (4.6~\mum), and the SiO fundamental (8~\mum).
Additionally, a complex set of narrow absorption features appears at
$\sim$3--4 \mum.  Most of these sources are K and M giants and
supergiants with C/O ratios less than unity, although there is one F
dwarf and 2 S stars.

\noindent{\bf 1.NC}  The 1.NC stars show several molecular
absorption bands indicative of a carbon-rich photosphere, including
narrow bands at $\sim$2.5~\mum\ (attributed to CO, CN, and C$_2$),
and 3.1~\mum\ (HCN and C$_2$H$_2$) and broad bands at $\sim$5~\mum\
(C$_3$, CO, and CN), 7--8~\mum\ (HCN, C$_2$H$_2$, and CS), and
14--15~\mum\ (HCN and C$_2$H$_2$)
\citep[e.g.][]{g78,g80,ato98a,ato99}.

\noindent{\bf 1.NE}  The 1.NE stars are emission line stars.
Numerous hydrogen recombination lines appear in emission.
Recombination lines from helium \citep{hwz00} or fine-structure
lines from, for example, [\ion{Fe}{2}] and [\ion{Ni}{2}] \citep{lam96} may also
be present.  In some sources, Balmer-like jumps for the infrared
series, such as the Humphreys jump 6--$\infty$ near 3.4~\mum, produce
discontinuities in the continuum \citep{hwz00}.

 \cite{hsp01} have classified  1.N and 1.NO sources in
more detail, reaching Level 3 for $\sim$40 sources.  They 
distinguish sub-classes of stars with (1) only strong H lines, 
(2) strong CO absorption and no SiO, (3) strong CO and SiO
absorption bands; and (4) strong CO and SiO features plus the \htwo\ 
bending mode feature.  The strength of the molecular features increase with
decreasing temperature and, consequently, later MK class.  They also find
that the strength of the infrared bands are well correlated with each
other.

\subsubsection{Group 2---Stars with Dust \label{sec.g2}}

\begin{figure}
\plotone{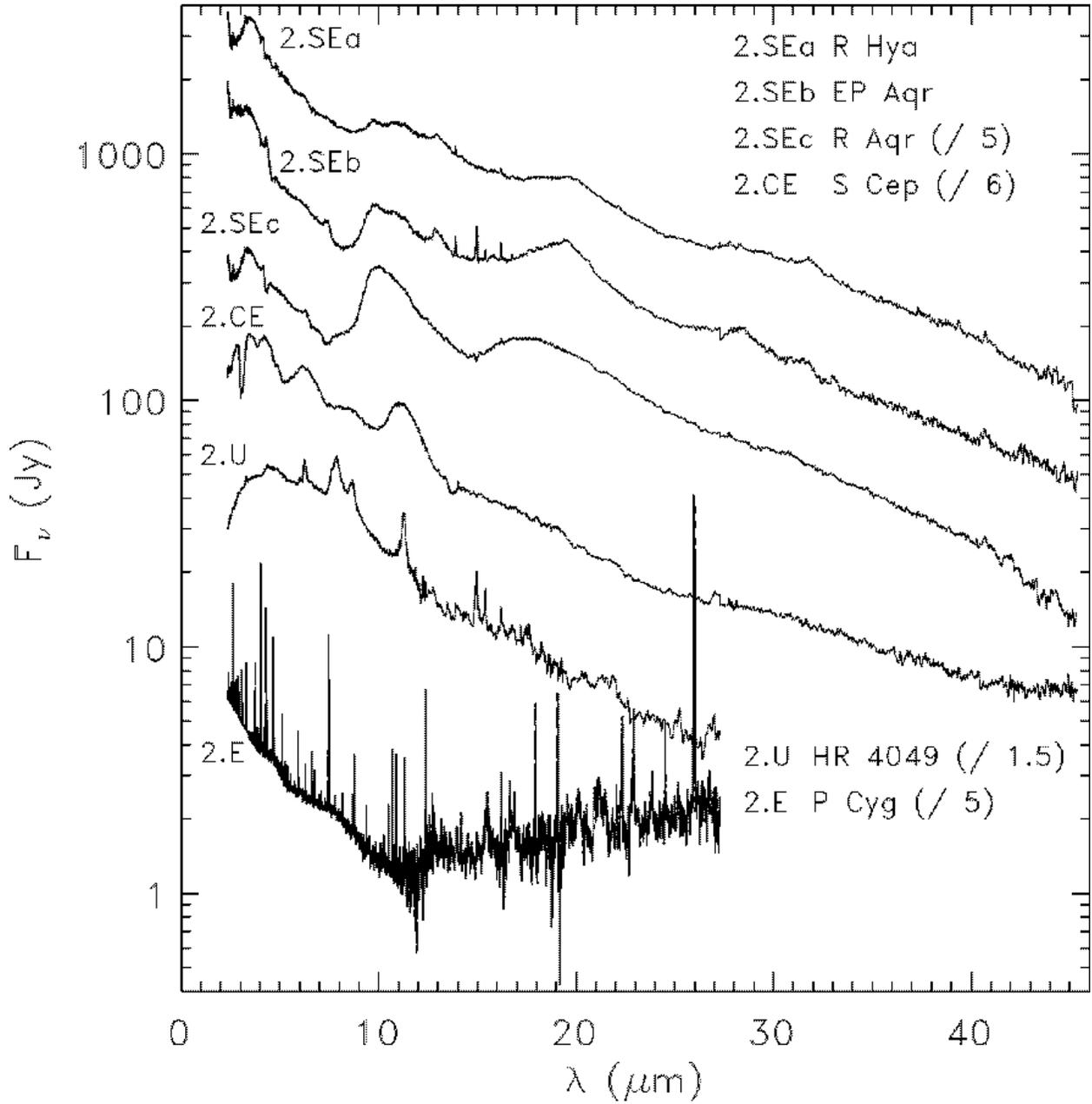}
\caption{Typical spectra in Group 2.  The 2.U and 2.E spectra are truncated
at 27.5 \mum\ (through Band 3D) due to poor signal-to-noise in Bands 3E and 4.
}
\label{fig.2}
\end{figure}

Group 2 includes sources with SEDs dominated by the stellar
photosphere but also influenced by dust emission (Fig.
\ref{fig.2}).  The nature of the spectral contribution from the dust
in the mid-infrared (typically $\sim$10--11~\mum) determines the
subgroup.  The dust properties are usually consistent with the
photospheric features in the near-infrared.  Most of the sources show
oxygen-rich dust emission (SE), and we have separated these spectra
into three subgroups based on the shape of the spectral emission
feature in the 10--12~\mum\ region analogous to classes defined by
LML and SP.

\noindent{\bf 2.SEa}  These spectra show a broad emission feature 
peaking $\sim$12~\mum.  The dust emission is usually weak, so the 
spectra resemble those in subgroup 1.NO.  The LML system classifies
these as ``broad'' spectra, and the SP system classifies them as
SE1--3.  This broad feature arises from amorphous alumina dust
\citep{ojw89,lmp00}.  A weak 20~\mum\ silicate feature is usually
present, as well as a complex of absorption bands $\sim$3~\mum\
(from ro-vibrational \htwo\ transitions and a broad, deep OH
transition).  Some sources show the well-known 13~\mum\ emission
feature, often associated with narrow CO$_2$ emission bands at
13.87, 14.97, and 16.28~\mum\ \citep{jfj98}.  Roughly 15\% of the SEa
sources have particularly weak dust emission. \cite{hsp01} note that one
of these, V Nor, has anomalous mid-infrared properties relative to 
its optical classification, probably the result of an unrecognized thin
dust shell.

\noindent{\bf 2.SEb}  The SP system describes these spectra as 
``structured silicate emission'' (SE3--6), while the LML system would
classify them as ``S'', ``3-component'', or ``Sil++''.  These spectra
have 10~\mum\ dust features due to amorphous silicates, but they also
show a secondary peak to the emission $\sim$11~\mum, and they often
have a 13~\mum\ feature as well (with associated CO$_2$ bands).  The
18--20~\mum\ feature tends to have a moderate strength, and the
\htwo\ feature at 3~\mum\ in 2.SEa and 1.NO sources is also present,
although with less influence from OH.  The spectral structure at 10
and 11~\mum\ may arise from a mixture of amorphous alumina or
silicate grains \citep{lmp00} or from optically thick but
geometrically thin shells of pure amorphous silicates \citep[where
the emission feature has begun to self absorb;][]{es01}.

\noindent{\bf 2.SEc}  These sources have strong silicate emission 
features with peaks at 10 and 18~\mum.  A few sources also show the
13~\mum\ feature.  The LML system classifies these as ``Sil'' or
``Sil+'' and the SP system describes them as ``classic silicate
spectra'' with SE indices of 6--8.  The photospheric absorption bands
shortward of 10~\mum\ are often complex.

\noindent{\bf 2.CE}  These spectra have a strong emission feature at
$\sim$11.5~\mum\ due to SiC dust emission.  Photospheric features
include bands at 2.5 and 3.1~\mum\ attributed to HCN and \ctht,
and at 4.3--6.0~\mum\ attributed to CO and C$_3$
\citep[e.g.][]{hlh98, jhl00}.  The complexity of the emission and
absorption features shortward of 10~\mum\ makes it difficult to
determine the continuum level in this wavelength region.

\noindent{\bf 2.C/SE}  These spectra have carbon-rich photospheric features
but the oxygen-rich silicate emission feature at 10--12 \mum. Two known
silicate carbon stars, V778 Cyg and W Cas, are tentatively joined in this
subgroup by RZ Peg.

\noindent{\bf 2.U}  The two sources in this category show  stellar 
photospheres with superimposed UIR emission features.  The photospheric 
spectrum for XX Oph resembles the 1.NO sources. The photosphere for HR 4049,
an unusual low-metallicity, high mass-loss, post-AGB star \citep[e.g.][and 
references therein]{vw95}, is unique in the SWS database.

\noindent{\bf 2.E}  These sources show emission lines on a
photospheric SED, with possible weak dust emission features in the
12--20~\mum\ range.  The exception is WR 147, which may have
silicate absorption in its spectrum \citep{mvc00}.

\noindent{\bf 2.M}  This subgroup includes miscellaneous spectra
which contained dust emission but could not be assigned to another
subgroup, primarily due to a poor signal/noise ratio in the
$\sim$10--12~\mum\ region. %OK?

\subsubsection{Group 3 \label{sec.g3}}

\begin{figure}
\plotone{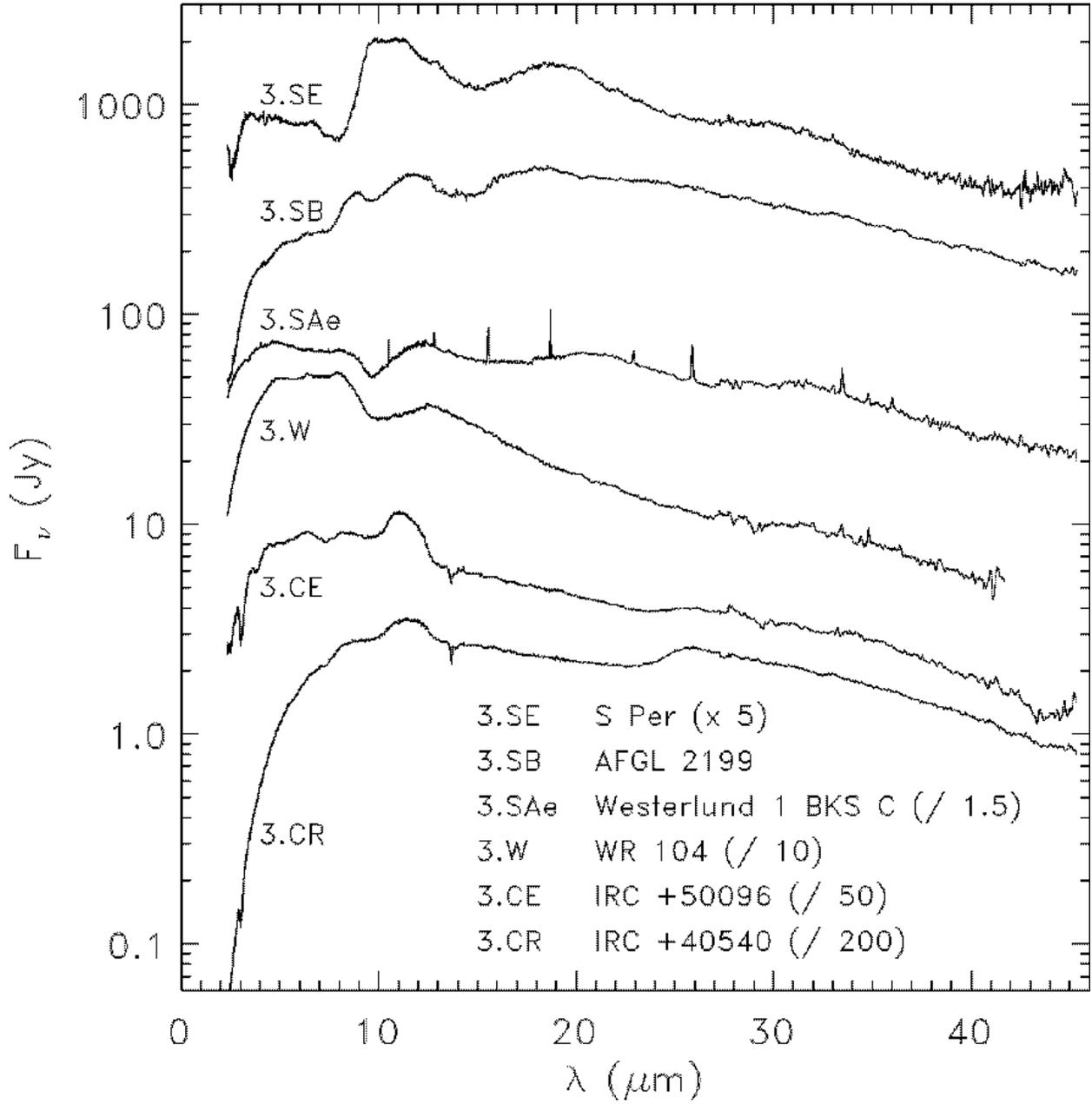}
\caption{Typical spectra in Group 3. 
}
\label{fig.3}
\end{figure}

Emission from warm dust dominates the SEDs of Group 3; this dust
emission usually arises from a circumstellar shell.  The spectra peak
shortward of $\sim20$ \mum, usually 10--15 \mum, but they show little
or no contribution from a stellar photosphere.  Like the previous
groups, the carbon and oxygen sequences are quite distinct, as Figure 
\ref{fig.3} illustrates.

\noindent{\bf 3.SE}  These sources show silicate emission at 10~\mum\ 
superimposed on the thermal continuum from the dust shell.  The dust
emission features resemble the classic silicate features in subgroup
2.SEc (at 10 and 20~\mum), but with no photospheric emission present
due to the optically thicker dust shell.  Three sources, all
symbiotic novae, show several forbidden emission lines (3.SEe), 
notably [\ion{Ne}{6}] at 7.65~\mum, [\ion{Ne}{5}] at 14.32 and 24.32~\mum, and
[\ion{O}{4}] at 25.89~\mum.  Three other sources have peculiar spectra
(3.SEp) with a weak or missing 20~\mum\ emission feature; two of
these are S stars.  The more typical 3.SE sources tend to be AGB
sources, OH/IR stars, or supergiants, although three of these 14 are
pre-main-sequence Ae or Be stars.

\noindent{\bf 3.SB}  These spectra arise from optically thick shells;
self absorption of the silicate dust has shifted the 10~\mum\ feature
closer to 11~\mum.  The SEDs peak at $\sim$18--19~\mum, and some
sources show crystalline silicate features longward of 30~\mum.  The
sources are associated with the AGB or OH/IR stars.

\noindent{\bf 3.SAe}  This unusual subgroup shows a 10~\mum\
absorption feature and bright emission lines from [\ion{S}{4}] at
10.5~\mum, [\ion{Ne}{2}] at 12.8~\mum, [\ion{S}{3}] at 15.6~\mum, 
[\ion{Fe}{3}] at
22.9~\mum, and [\ion{S}{3}] at 33.5~\mum.  The 20~\mum\ silicate
emission feature is more rounded than in the SE and SB spectra and
appears at a slightly longer wavelength.  Both sources are young or
pre-main-sequence Be stars.

\noindent{\bf 3.CE}  These spectra resemble the 2.CE subgroup,
showing an emission feature $\sim$11.5~\mum\ from SiC, but dust
absorption obscures the photospheric absorption features from
molecular bands which dominate the near-infrared wavelengths of the
2.CE and 1.NC spectra.  Most of the sources show a narrow and often
deep \ctht\ absorption band at 13.7~\mum.

\noindent{\bf 3.CR}  These sources are cooler analogs of the 3.CE
sources.  The 11.5~\mum\ SiC emission feature still dominates and
the \ctht\ absorption band at 13.7~\mum\ is still prominent, but
other emission features also appear, usually in the 26--30~\mum\
region and sometimes at 8~\mum.  \mbox{IRC +10216} is the brightest
of these sources; radiative transfer modeling of its spectrum
suggests that amorphous carbon dominates the SiC dust component
\citep[$\sim$90--95\%; e.g.][]{mr87,se95}.  The optical efficiency
of amorphous carbon follows a $\lambda^{-1}$ relation in the
mid-infrared, mimicking a blackbody of lower temperature than the
actual dust temperature \citep{mr87}.

\noindent{\bf 3.W}  The spectra of these sources peak at
$\sim$6--8~\mum.  Most show strong silicate absorption features at
10~\mum\ similar to the 4.SA feature.  Except for the 10~\mum\
feature, the spectra are nearly featureless and have little
similarity to sources in any other subgroup.  All members are either
Wolf-Rayet or R Corona Borealis stars.

\subsubsection{Group 4 \label{sec.g4}}

\begin{figure}
\plotone{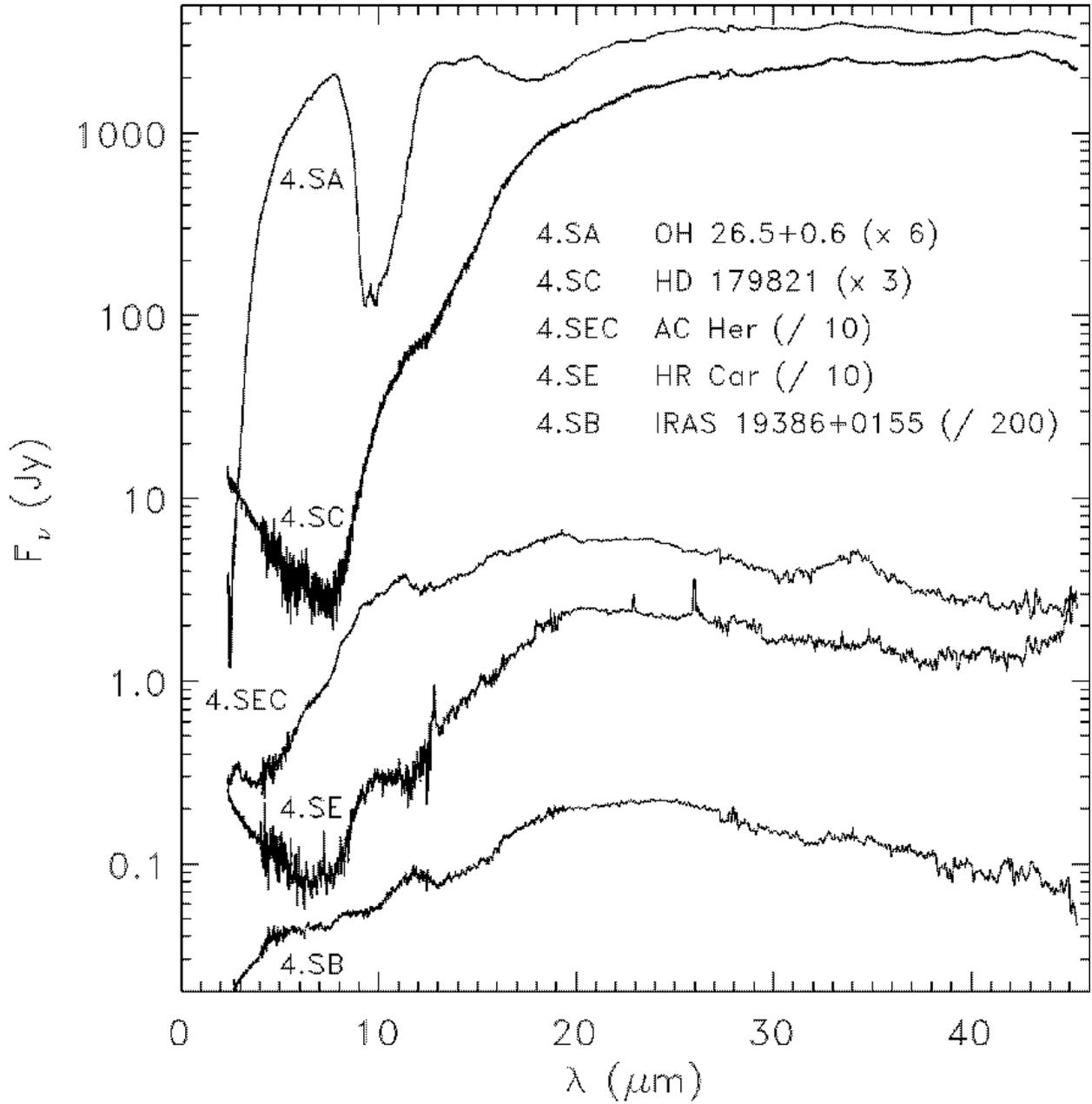}
\caption{Typical spectra for Group 4. (a) oxygen-rich sources, arranged 
in a possible evolutionary sequence (\S \protect{\ref{sec.orich}}). 
%(b) Carbon rich sources.  The spectrum for 4.CR
%is truncated at lower wavelengths due to poor signal-to-noise.
%(c) The spectra for 4.PU and 4.F
%are truncated at lower wavelengths due to poor signal-to-noise.
}
\label{fig.4a}
\end{figure}

\begin{figure}
\figurenum{4}
\plotone{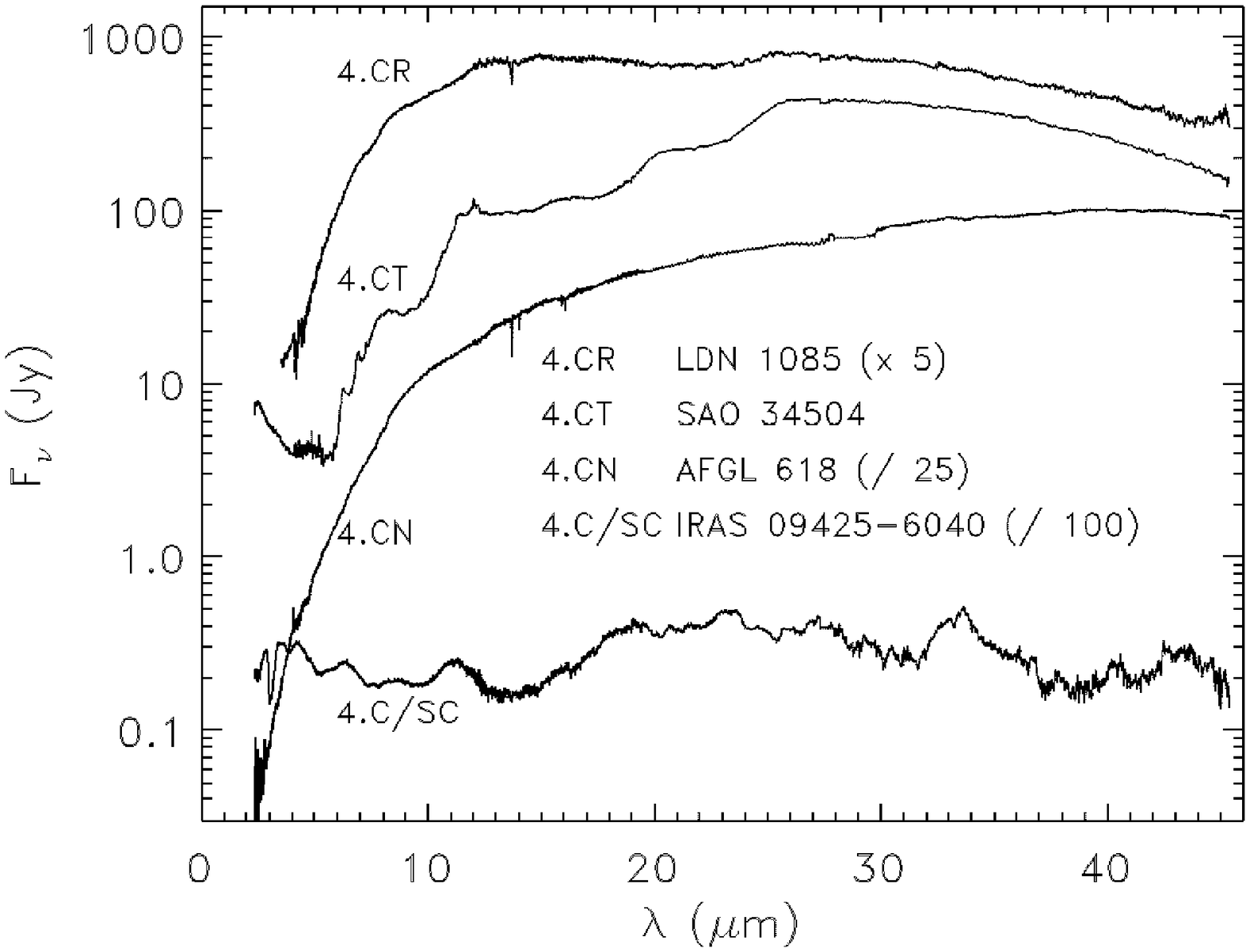}
\caption{Group 4, cont'd. (b) Carbon rich sources.  The spectrum for 4.CR
is truncated at lower wavelengths due to poor signal-to-noise.
}
\label{fig.4b}
\end{figure}

\begin{figure}
\figurenum{4}
\plotone{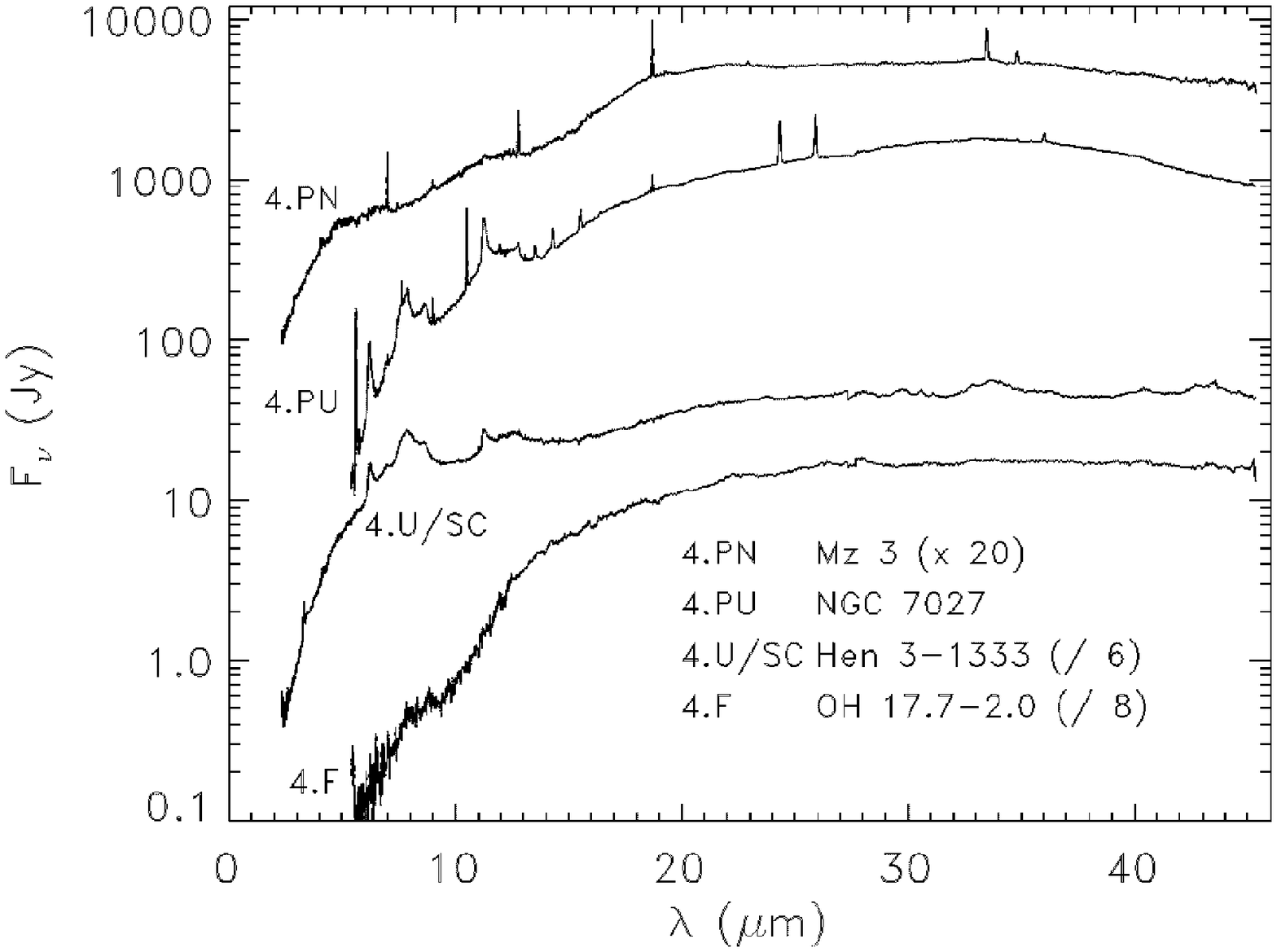}
\caption{Group 4, cont'd. (c) The spectra for 4.PU and 4.F
are truncated at lower wavelengths due to poor signal-to-noise.
}
\label{fig.4c}
\end{figure}

Dust emission dominates the SEDs of Group 4, and the dust temperature
is cooler than in Group 3, with the spectra peaking at wavelengths
between $\sim$20 and $\sim$40~\mum.  The photospheric contribution is
generally negligible. Several of the subgroups in Group 4 are
analogous to those in Group 3, but with significantly cooler dust.
As in the warmer groups, the carbon and oxygen sequences are quite
distinct.  The carbon-rich spectra continue in relatively tight
groups whereas the oxygen-rich dust spectra form a rather
heterogeneous group.  Finding distinct and self-consistent subgroups
for these spectra has proven difficult, and finding subgroups
populated by uniform samples has proven impossible. Figure 4 
 shows sample spectra from Group 4.

\noindent{\bf 4.SE}  These sources show an emission feature from 
amorphous silicates at 10~\mum\ superimposed on emission from a cool
dust shell.  As in subgroup 3.SE, the contrast varies significantly
from one source to the next.  Most of the SEDs peak around
20--25~\mum, although in three sources the peak is near 30~\mum.  Of
all the subgroups in Group 4, this is the most difficult to
characterize, due in part to the low signal/noise ratio of many of
the spectra.  The specific shapes of the SED and the 10~\mum\ silicate
feature differ among the sources.  Some spectra, often with bluer
SEDs, show forbidden emission lines (the specific transitions vary
substantially from source to source).  Most 4.SE sources
 have optical spectral classes of Be, Ae, or Fe, or
are described as PNe.  While more than half of the sources are 
post-main-sequence, several are Herbig Ae/Be stars or related 
pre-main-sequence objects.

\noindent{\bf 4.SEC} These sources show prominent crystalline
silicates in at least two of three positions:  $\sim$11, 23, and
33~\mum.  The 11~\mum\ emission feature may be weak, in which case
the usual amorphous feature at 10~\mum\ appears to be broadened, or
it may dominate, producing a strong, sharp peak at 11~\mum.  Additional
crystalline silicate features may also be present at 19 and 43~\mum.
Like 4.SE, this subgroup includes a heterogeneous collection of
sources, but only three of the 11 objects are clearly identified as
pre-main-sequence.  The rest tend to be young PNe or proto planetary
nebulae (PPNe); the sample also includes one Mira variable and the
hypergiant IRC +10420.

\noindent{\bf 4.SB}  The 10~\mum\ silicate feature is in
self-absorption, sometimes strongly, and emission features from
crystalline silicates may be present at 33, 40, and/or 43~\mum,
though not strongly.  The majority of sources are PNe or PPNe,
although one may be a pre-main-sequence Ae star, another is a Be
variable, and a third is an AGB source.

\noindent{\bf 4.SA}  This subgroup exhibits silicate absorption at
10~\mum\ and sometimes also at 18~\mum.  Most sources show emission
features from crystalline silicates at 33, 40, and 43~\mum.  Stronger 
absorption at 10~\mum\ usually occurs with stronger emission
from crystalline silicates, especially at 33~\mum.  The deepest
10~\mum\ absorption features arise from OH/IR stars.  Other sources
in this subgroup include PPNe and PNe.  Two of the three bluest
sources are more difficult to characterize and may be
pre-main-sequence.

\noindent{\bf 4.SC}  All sources in this subgroup show crystalline
silicate emission features at 33 and 43~\mum.  Many also show
crystalline silicate features at 23 and 40~\mum.  The source types
are somewhat heterogeneous, although PPNe and PNe (especially young
PNe associated with Be central stars) dominate.  Other sources
include Wolf Rayet stars, OH/IR stars, and one source identified as a
pre-main-sequence G star (DG Tau).

\noindent{\bf 4.F}   The SEDs of sources in this subgroup basically
have no features (greater than the noise), with no silicate emission
or absorption at 10~\mum\ or any crystalline silicate emission at
longer wavelengths.  Three sources do show UIR emission, and one has
several non-silicate absorption features due to ices.  Most sources
are PNe or OH/IR stars, except for the source with ice absorption,
which is a YSO (R CrA [TS84] IRS 2).

\noindent{\bf 4.CR}  This subgroup continues the carbon-rich dust
sequence (2.CE---3.CE---3.CR) to cooler shells.  The SEDs peak
$\sim$28~\mum, and are broad and nearly featureless, except for the
26--30~\mum\ emission feature and the \ctht\ absorption feature at
13.7~\mum.  Extreme carbon stars and carbon-rich PPN candidates
dominate the source types.

\noindent{\bf 4.CT}  These sources have SEDs with a step-like 
appearance produced by emission features on a steadily rising red
continuum at 8, 11.5, 21, and 26--30~\mum.  The 21~\mum\ can be
prominent, and the SEDs peak $\sim$30~\mum.  Unlike the other dusty
carbon-rich sources, they do not have the \ctht\ absorption feature
at 13.7~\mum.  Sources tend to be F or G supergiants sometimes
identified as PPN candidates.

\noindent{\bf 4.CN}  These sources show features such as the \ctht\ 
13.7~\mum\ absorption or the 11.5 \mum\ emission feature, which
indicate they are carbon-rich.  The SED peaks $\sim$40~\mum.  All of
the sources are identified as PPNe, and this subgroup includes the
well-known carbon-rich bipolar nebulae AFGL 618 (the Westbrook
Nebula) and AFGL 2688 (the Cygnus Egg).  The ``N'' designation stands
for ``nebula.''

\noindent{\bf 4.C/SC} The one source in this subgroup (IRAS
09425$-$6040) has an unusual spectrum, showing carbon-rich molecular
absorption bands in the near infrared and SiC emission
$\sim$11.5~\mum\ as seen in 2.CE spectra as well as strong
crystalline silicate emission features at 33, 40, and 43~\mum.
\cite{myw01} suggest that IRAS 09425$-$6040 may be in transition to a
Red Rectangle-like object (see subgroup 4.U/SC below).  Normally, a
unique spectrum would belong in a miscellaneous subgroup, but the
relation of this spectrum to the more numerous U/SC subgroup suggests
that more of these sources may be discovered in future observations.

\noindent{\bf 4.U/SC}  The sources in this subgroup combine strong
UIR features (at 6.2, 7.7--7.9, 8.6, and 11.2~\mum) and strong
crystalline silicate emission features (at 33, 40, and 43~\mum).
The 33~\mum\ feature can be quite prominent, and in the bluer
sources, can be accompanied by a 23~\mum\ emission feature also due
to crystalline silicates.  Most spectra show a possible emission
feature $\sim$28.5~\mum\, but the poor quality of Band 3E makes this
identification problematic.  All of the sources are PPN or PN, with
the exception of a single Herbig Ae/Be star (HD 100546).

\noindent{\bf 4.PN}   The dominant spectral feature in this subgroup
is the presence of  strong fine-structure lines superimposed on a
SED which peaks in the vicinity of 30~\mum.  The line-to-continuum
ratio can be 5 or greater in some instances.  All show, at a minimum,
[\ion{Ne}{2}] at 12.8~\mum\ and [\ion{S}{3}] at 18.7 and 33.5~\mum.  Other 
common lines include [\ion{Ar}{2}] at 6.99~\mum, [\ion{Ar}{3}] at 8.99~\mum, 
[\ion{S}{4}] at 10.5~\mum, [\ion{Ne}{3}] at 15.6 and 36.0~\mum, and 
[\ion{Si}{2}] at 34.8~\mum, as well as Br $\alpha$ and $\beta$.  Additional 
detected lines include [\ion{Ne}{5}] at 14.3 and 24.3~\mum, [\ion{Ne}{6}] at 
7.65~\mum, [\ion{Ar}{5}] at 7.90 and 13.1~\mum, [\ion{Ar}{6}] at 4.53~\mum, 
[\ion{O}{4}] at 25.9~\mum, [\ion{Mg}{4}] at 4.49~\mum, and [\ion{Mg}{5}] at 
5.61 and 13.5~\mum.  Some sources also show weak crystalline silicate 
features, especially at 33~\mum. All but one source are planetary nebulae; 
the exception, IRAS 05341+0852, is a PPN-candidate.

\noindent{\bf 4.PU}   Similar to 4.PN, these sources show strong 
UIR features in addition to the fine-structure lines.  BD +30 3639 shows
crystalline silicate emission at 33~\mum.  Most are planetary
nebulae, including one PPN candidate.  Three sources with fewer,
weaker emission lines than the typical PU spectrum are noted as
peculiar with the ``p'' suffix; otherwise their SEDs and UIR features
resemble the other members closely.

\noindent{\bf 4.M} Each of the four objects in this subgroup is unique. 
$\eta$ Car could be described as the prototypically strange spectrum
at all wavelengths.  Classification of its SWS data is further
complicated by the saturation (and automatic flagging) of most of
Band 3, the spectral region upon which much of the subgrouping in
Group 4 is based.  AG Car combines a Group 1 spectrum (1.NE) in the
near-infrared with a Group 4 spectrum (possibly 4.PUp) at longer
wavelengths.  Only a few other objects show this combination of hot
photospheric emission  with very cool dust. 
%(BC Cyg, HR Car, and AFGL 4106). HD 179821? 
IRAS 21282+5050 has very strong UIR features, most similar
to those in the Red Rectangle (HD 44179, 4.U/SC), but has no crystalline
silicate emission and a significantly bluer SED than members of the
4.U/SC subgroup.  HD 169142 is somewhat similar to the 4.U/SC or PU
groups in terms of its UIR emission and SED, but has no evidence for
crystalline silicates or emission lines in its admittedly weak, noisy
spectrum.

\subsubsection{Group 5 \label{sec.g5}}

\begin{figure}
\plotone{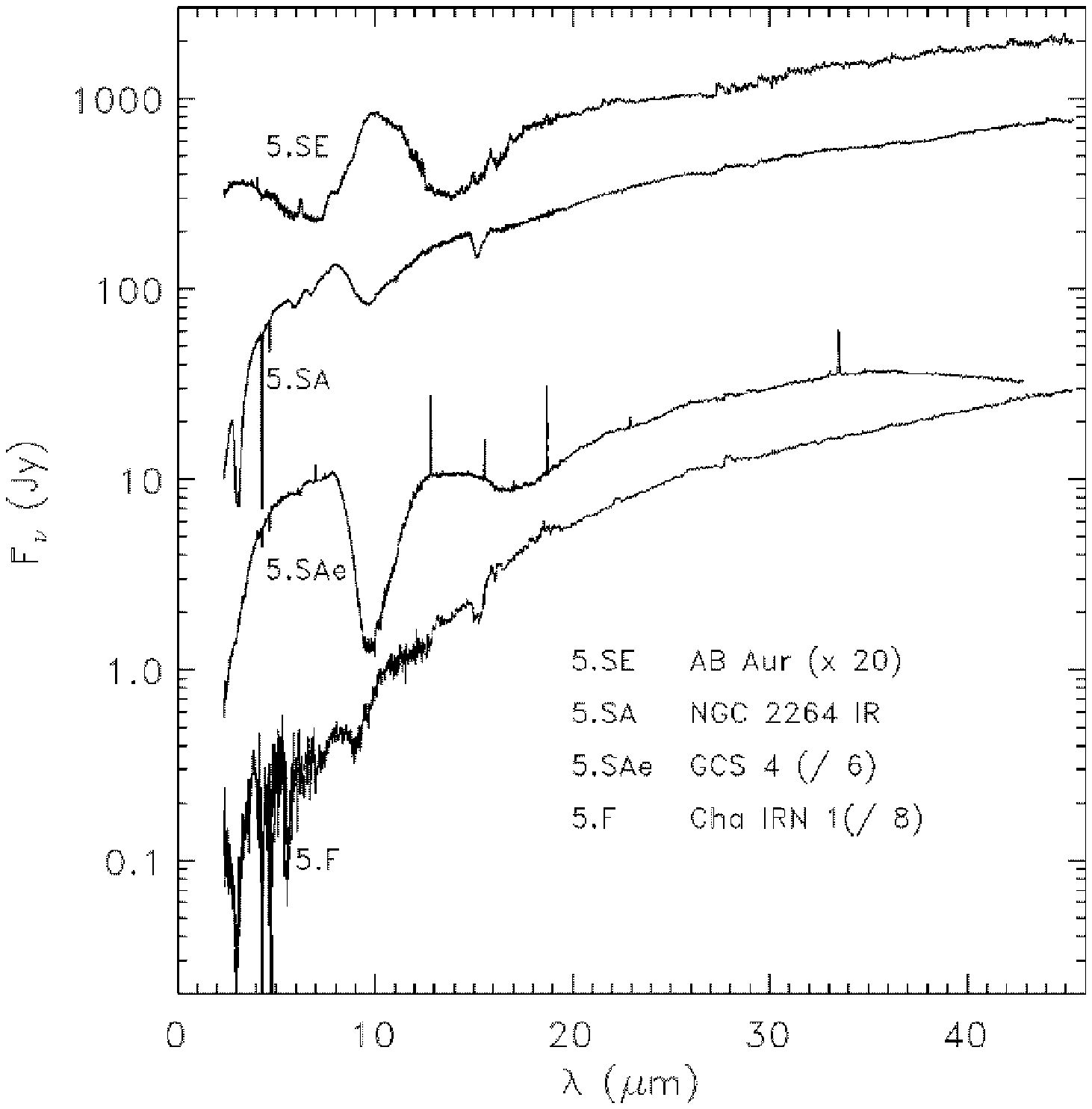}
\caption{(a)Typical spectra in Group 5. %(b) In 5.U, Band 2A is omitted due to
%poor signal-to-noise. Likewise, the spectra for 5.E and 5.PN are truncated
%at shorter wavelengths. 
}
\label{fig.5a}
\end{figure}

\begin{figure}
\figurenum{5}
\plotone{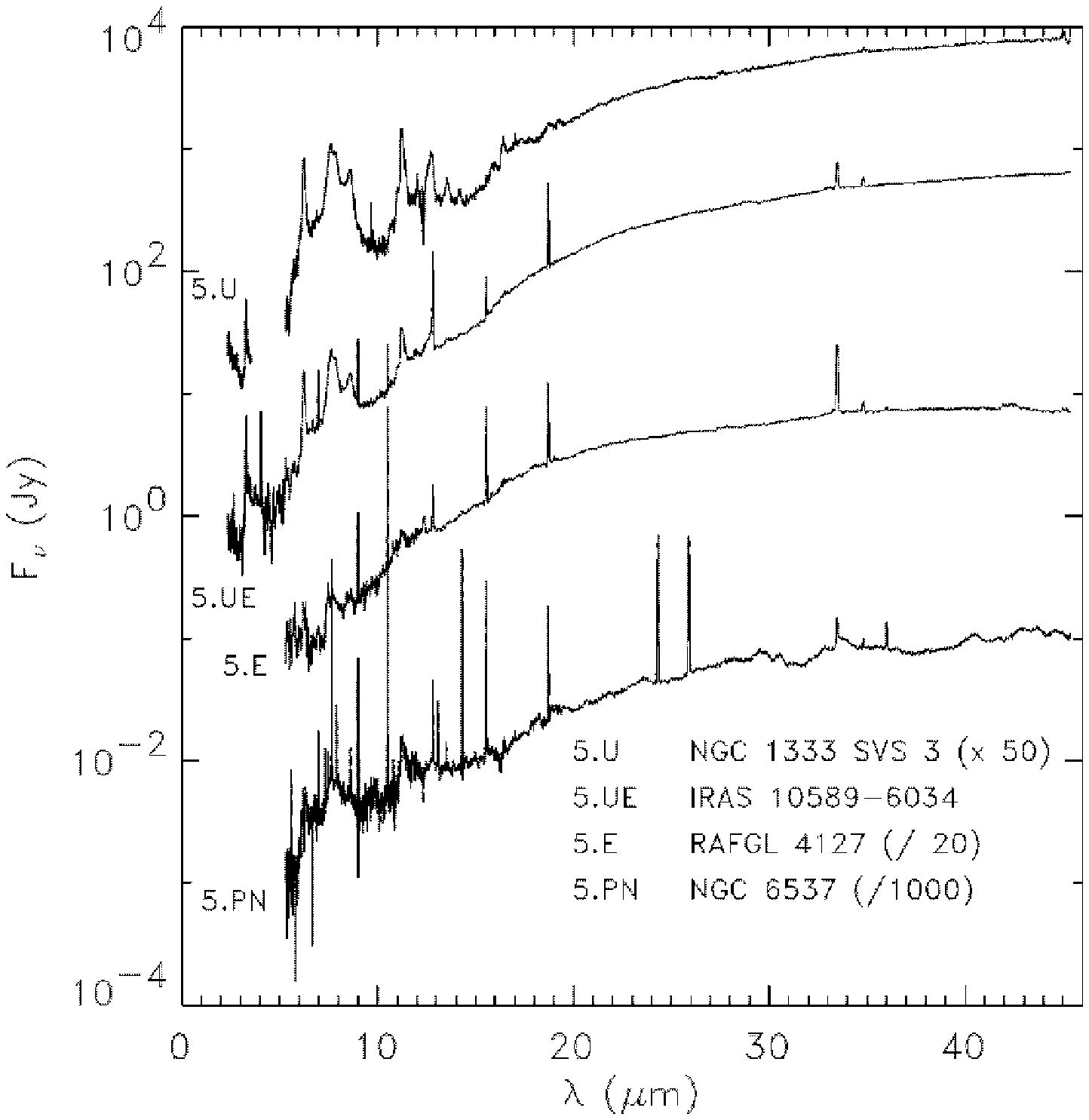}
\caption{Group 5, cont'd. (b) In 5.U, Band 2A is omitted due to
poor signal-to-noise. Likewise, the spectra for 5.E and 5.PN are truncated
at shorter wavelengths.
}
\label{fig.5b}
\end{figure}

Objects in Group 5, whose SEDs are still rising through the end of Band 4,
have the coolest dust emission in the database.
The subgroups trace the presence of silicate emission or absorption, 
narrow emission lines, UIR features, and absorption features. Figure 5 shows
sample spectra for Group 5.

\noindent{\bf 5.SE} These sources show broad silicate emission
features at $\lambda\sim9$--11 \mum.  One (AB Aur) also shows UIR
emission features.  All but one are young, Herbig Ae/Be (or Fe)
stars.  The single evolved source, HD 101584, a PPN-candidate, could be
a cooler version of the 4.SE sources or it may actually be a young object
misclassified as old.

\noindent{\bf 5.SA} Sources in this subgroup show a broad silicate 
absorption feature at $\lambda\sim9$--11 \mum.  Other absorption
features often present include bands from CO$_2$, CO, and \htwo.  A few also
show UIR emission features or weak atomic fine-structure lines
([\ion{Ne}{2}], [\ion{S}{3}], or [\ion{Si}{2}]).  Almost all sources are 
YSOs or in star forming regions. The six (out of 50) which are not
YSOs are probably OH/IR stars. Four of the sources in this class
with emission lines (5.SAe) are Galactic center objects.

\noindent{\bf 5.F}  These sources show no strong features
superimposed on a SED which rises steadily to the red.  Some sources
in this class may be better placed in other classes, but because the
red end of the spectrum is so strong, any structure at
$\lambda\la15$~\mum\ is not visible on the self-scaled plots used for
classifying.  Three of the sources are evolved; %(need to check)
 the rest are young.

\noindent{\bf 5.U}  These sources have moderate to strong UIR
features but no atomic fine-structure lines.  Only one source is
considered evolved (Wray 15-543, thought to be a PPN-candidate); the
rest are young.

\noindent{\bf 5.UE}   These sources have moderate to strong UIR 
features and strong atomic fine-structure lines.  The majority of
the sources are young; a few are thought to be
evolved (PNe).

\noindent{\bf 5.E} These sources have strong atomic fine-structure 
lines but little or no UIR emission.  The composition of this
subgroup is similar to 5.UE:  mostly pre-main-sequence with a few
PNe.

\noindent{\bf 5.PN}  These sources have very strong, numerous
atomic fine-structure lines.  Crystalline silicate emission is often 
present in the $\lambda\sim$30--45~\mum\ range, and at least two show
UIR emission.  All sources are evolved (PNe).

\noindent{\bf 5.M} The weak signal and poor signal-to-noise ratio of
these spectra hide any identifying features which would help to
place them in a different subgroup.

\section{Discussion \label{sec.disc}}

\subsection{Calibration Issues and the Classifications \label{sec.caliss}}

As mentioned in \S \ref{sec.data}, the browse products used to
classify most of the spectra did not fully correct for flux
discontinuities between bands.  The most challenging normalization
problems occur between Bands 2C and 3A, at $\lambda\sim12$~\mum, and
between Bands 3D, 3E, and 4, at $\lambda\sim$26--30~\mum.   We 
discuss them briefly here to the extent that they influence the
classification effort.

Spectra in Group 2 are most sensitive to discontinuities and memory
effects near 12~\mum, because the shape of the emission and
absorption features in the 10--12~\mum\ region serve as the primary
features for classification into the subgroups.  If the flux discontinuity 
is simply related to a gain difference between bands, normalization during
reprocessing (if needed at all) would simply scale Band 3A to match
2C without changing the basic shape of any features present.  If the
discontinuity results from memory effects, however, it is more
problematic.  Even with the {\it dynadark} correction and
normalization some error may remain in the shape of the spectrum.
Fortunately, this problem does not compromise the classification of a
spectrum as oxygen or carbon-rich, although it might cause a spectrum
to be (mis)classified as 2.SEa instead of 2.SEb, for instance.

Normalization of Bands 3D, 3E, and
4 is complicated by a light leak and by the unreliability of 3E.
Most spectra show a smooth shape with a roughly constant slope from Band 3D
through Band 3E and into Band 4, which allows a straightforward
normalization of these bands to each other.  However, spectra with
structure near 26~\mum\ present more of a problem, since the changing
slope of the spectrum makes extrapolation across Band 3E difficult.
This problem affects the carbon-rich sources in Groups 3 and 4 most
significantly and  limits our confidence in the shape of the
emission feature in the 26--30~\mum\ region.  In the browse product
spectra produced from OLP 7.1, the normalization of the segments
makes the 26--30~\mum\ feature appear narrow and peaked around
$\sim$25--29~\mum.  Applying our normalization algorithm to data
in OLP 10.0 broadens the feature to $\sim$25--34~\mum.  The
literature tends to refer to this feature as the 30~\mum\ emission
feature, possibly attributable to MgS \citep{gm85,beg94}.  With the
current uncertainties in calibration, we are unable to definitively
address this issue.

To date, no model has been developed to correct the memory effects in 
Band 4.  The entire shape of Band 4 can be compromised, and, in terms of the
spectral classification, this influences whether a spectrum is classified in
Group 4 or 5.  For example, a spectrum could be misclassified, as a
4.PN instead of a 5.PN because the spectrum appears to have turned
over in Band 4 when it is actually still climbing.  The memory effect
in Band 4 can also influence our ability to recognize crystalline
silicate features, especially at 40 and 43~\mum.  These features
could be washed out when the two scan directions are combined
because of the difference in flux levels between them.  As with the
other issues raised here, the Band 4 memory effect should have a limited
impact on the classifications.

Despite these issues, the basic classification scheme and the
grouping of the spectra should prove robust.  The movement of a few
spectra from CR to CE or from Group 4 to Group 5 will not change the
overall nature of the database or the existence of any of the
evolutionary patterns discovered therein (\S \ref{sec.co}--\ref{sec.crich}).

\subsection{Comparison With \iras\ Classifications \label{sec.irascomp}}

Although we are dealing with a non-uniform database 
(\S \ref{sec.seleff}), we can still
compare our classifications with the LRS classes \citep{lrs88}
and the classes of \citet[][hereafter KVB]{kvb97}.
Only the subset of SWS sources with LRS classifications (379 sources) or 
KVB (567 sources) classifications can be 
considered, so the numbers quoted below will not
be the same as those given for each subgroup in Table \ref{tab.cross1-2b}.
Also, recall that for LRS class $1n$, $n=2\beta$ where $\beta$ is the 
spectra index:

\begin{eqnarray}
F_{\lambda} & \propto & \lambda^{-\beta}.
\end{eqnarray}

\noindent Thus, when $\beta=4$, the spectrum behaves as a pure Rayleigh-Jeans 
tail and is in LRS class  18.  Sources with low-contrast dust mimic lower 
spectral indices and receive lower LRS characterizations.  For example, LML 
and SP showed that many sources in LRS classes  13--16 show low-contrast 
alumina-rich dust in their spectra.

\subsubsection{Similarities \label{sec.sim}} 

Group 1, the dust-free stars, corresponds well to the LRS classes 17--19. 
Of the 60 objects in Group 1 with LRS classifications, 54 are in LRS classes
17--19, with 39 in class 18. Of the 2.SEa sources, with low-contrast dust,
81\% of the 53 sources are in LRS classes 13--16, as expected. Similarly, 
the oxygen-rich  dust sequence, described below in \S \ref{sec.orich}, should 
begin in the  $2n$ range and progress to the $3n$ range, where $2n$ 
corresponds to silicate emission and $3n$ to 
silicate absorption. Nearly 90\% of the 67 objects in subgroups 2.SEb,
2.SEc, and 3.SE have LRS classes $2n$. The sources in 3.SB, the
self-absorbed subgroup, are split between $2n$ and $3n$, and 11 of 12
sources in 4.SA are $3n$ or $7n$ (recall that $7n$ is the red counterpart
of $3n$). In the carbon-rich sources, 31 of 37 sources
in subgroups 2.CE, 3.CE, and 3.CT have LRS=$4n$, the carbon-rich LRS
class. Only about a third (14 of 41) of the PNe subgroups 4.PN, 4.PU, and 
5.PN, have LRS classifications, but those that do tend (11 of 14) to be 
$9n$, that is, red objects with emission lines but no detected 11.3 
\mum\ UIR feature. For the young, red sources in Groups 4 and 5, even 
fewer, $\sim25$\%, have LRS classifications, so small numbers make valid 
comparisons problematic.  Still, most of those with LRS data in our SA or 
SE subgroups do have silicate absorption or emission LRS classifications.

A comparison of our classifications with those of KVB shows 
comparable similarities. For example, 37 of the 40 sources with their
class C, for carbon-rich, are in one of our carbon-rich subgroups 
(mostly 2.CE). More than 80\% of their A (10 \mum\ absorption) sources 
are in our SA or SB subgroups and more than 90\% of their E (10 \mum\ 
emission) sources are in our silicate emission subgroups. Almost 90\%
of their S (stellar) sources are in Group 1, our naked star category.

\subsubsection{Distinctions \label{sec.diff}}

\begin{figure}
\plotone{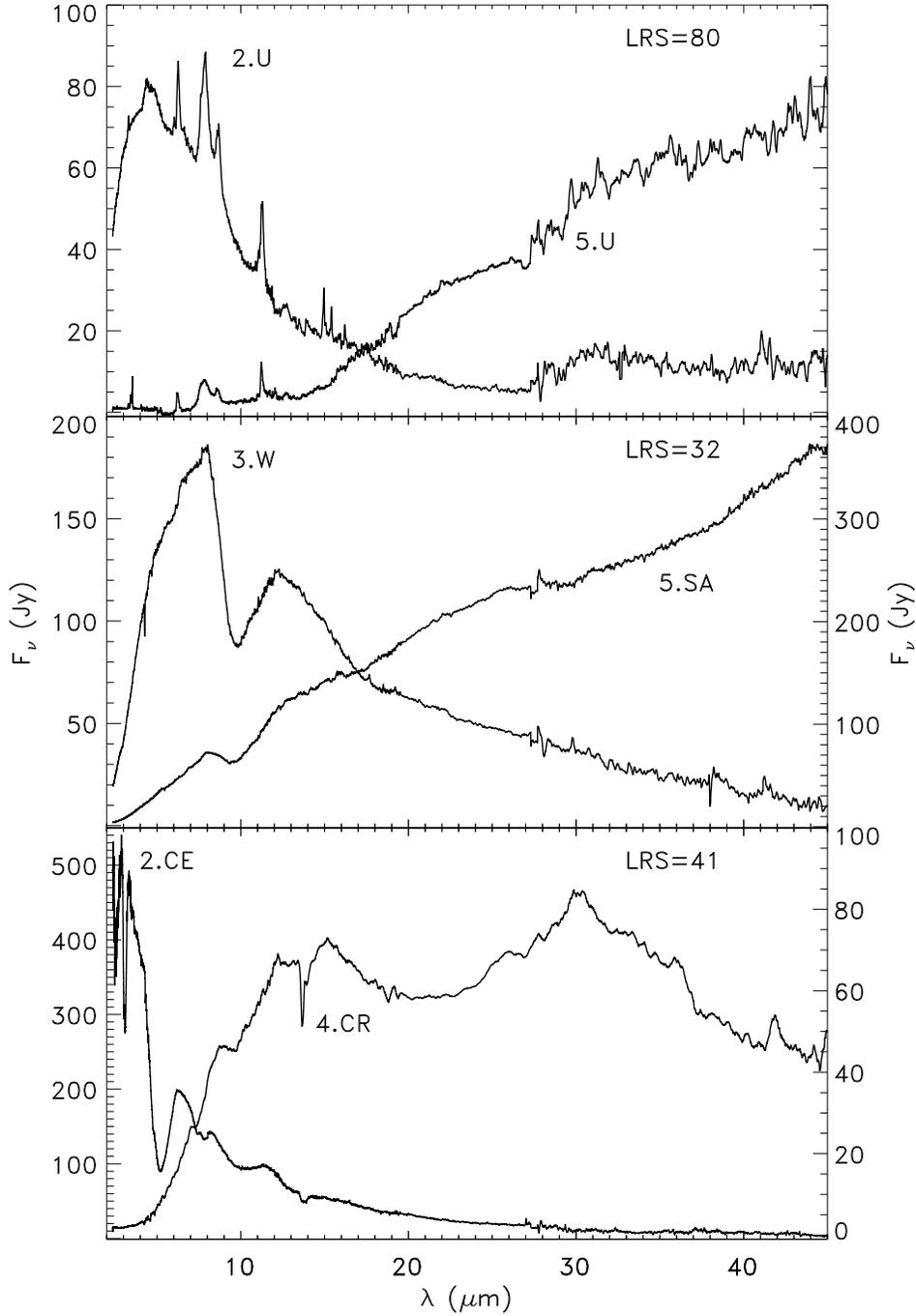}
\caption{Comparison of spectra with the same LRS class but different KSPW
classes. Top: LRS=80 (UIR emission): HR 4049 2.U and HD 97048 5.U (smoothed).
Middle: LRS=32 (blue SED + silicate absorption): WR 112 3.W and V645 Cyg 5.SA.
Bottom: LRS=41 (C-rich): RY Dra 2.CE and IRAS 22303+5950 4.CR.
}
\label{fig.lrscat}
\end{figure}

While the overall correspondence betweeen our classifications and
those from LRS-based schemes is reasonable, there are a number of 
important differences.  For instance, 
misidentification of UIR features as silicate absorption occurred in 
the LRS classifications due to the low spectral resolution and bandwidth.
This is largely avoided in the SWS database because of the higher
spectral resolution and especially the expanded bandwidth. The extended
wavelength coverage allows confirmation of suspected 7--11 \mum\ UIR features 
with those at 6.2 and 3.3 \mum\ which were outside the LRS range.

It was mentioned above that most of the 4.SA and 5.SA sources  (24 of 27)
were in LRS classes corresponding to silicate absorption. However, 17
of those sources were $3n$, with ostensibly blue SEDs. Characterizations
of $7n$ would have been more correct, but the short wavelength cutoff of
only 7 \mum\ presumably prevented an accurate assessment of the overall SED.
In the 5.UE group,  25 sources had red LRS characterizations, but less 
than half (10) were $8n$, the UIR+emission line class. Lack of sensitivity
of the LRS precluded the detection of UIRs in some sources, while the limited 
spectral range caused the confusion of UIRs and silicate absorption in 
others.

Examination of the carbon-rich classes and ostensibly carbon-rich objects
further illustrates  the limitations of the old LRS classifications 
when dealing with SWS data. One source, AFGL 2287, was classified as 
carbon-rich in all three LRS-based schemes but is classified by us as 
self-absorbed silicate emission (3.SBp). In the limited spectral range 
of the LRS data, self-absorbed silicates can appear similar to carbon-rich 
spectra \citep{wc88}. However, with the SWS, 
AFGL 2287 can be seen to have none of the other features typical of 
carbon-rich sources such as the absorption features at 13.7 \mum\ or 
3 \mum.  Seven other sources 
were classified as carbon-rich in the LRS atlas but as oxygen-rich by 
AutoClass and KVB. With the high-quality data from SWS, we can confirm 
that they are indeed oxygen-rich\footnote{They are ST Her (2.SEa), FI Lyr 
(2.SEa), Z Cas (2.SEap), $\pi^1$ Gru (2.SEa), AD Per (2.SEa), AFGL 2199 
(3.SB), and AFGL 1992 (3.SB).}.

As an example of a discrepancy in the opposite sense, only one of our 4.CR 
sources has a carbon-rich LRS class, while most  of the rest (8 of 11) are 
classed as 21--23, low-contrast silicate emission. The AI classifications
also mistook most  of 4.CR (9 of 10) for oxygen-rich ($\zeta4$) because of the 
limited spectral range on which the classifications were based.
Of the 31 objects observed with SWS which SIMBAD lists as carbon stars, 
only two (FI Lyr and CIT 11) are in non-carbon KSPW subgroups; four more 
are in Group 7 or flagged, so 25 of 27 agree with SIMBAD. The
LRS scheme, on the other hand, has only 16 of 24 sources with $4n$ or 04
designations. Again, the superior sensitivity  and spectral range of SWS
enabled the proper classification of these sources as carbon-rich.

The LRS  $4n$ classes base the second digit on the strength of the 
11.5 \mum\ silicon carbide feature. However, nothing in the LRS
classification indicates the shape of the underlying continuum for 
carbon-rich objects. 
Thus, sources as dissimilar as RY Dra (2.CE) and IRAS 22303+5950 (4.CR)
possess the same LRS classification 41 because both have weak 11.5 \mum\
features (Fig. \ref{fig.lrscat}). LRS class 44 contains both 
W Ori  (2.CE) and IRC +50096 (3.CE) despite their distinctly 
different underlying 
SEDs.  There simply are {\em no} appropriate categories in the LRS scheme in
which to place {\em any} of the carbon-rich sources in our Groups 3 and 4
without loss of significant information about the spectra. 

Similarly,
placing all of our Group 1 naked stars into $1n$ would also lose important 
information about the photospheric chemistry (1.NO vs. 1.NC) or the presence
of emission lines (1.NE). Other KSPW classes with no good LRS counterparts
include 2.E, 2.U, 2.C/SE, 4.SC, 4.U/SC, 4.SEC, and 6.

Other LRS classes also   mis-matched sources, as Figure \ref{fig.lrscat} 
shows. These disparate sources were placed in the same 
LRS classes because of the limited spectral range of the LRS
and small number of features considered in that classification scheme.
Table \ref{tab.irascomp} compares the KSPW classes to the LRS classes.
Column 2 lists the KSPW classes that are well-matched to an LRS class as
defined in the \citet{lrs88},
for example 1.N and the $1n$ class. Column 3 lists our groups that could
be placed in an LRS class, but only with information lost, such as 2.C/SE
in $2n$ or 4.U/SC in $8n$. The last column lists the other KSPW groups in
which the LRS classes actually appear but are not well-suited to each other.
Application of the LRS scheme to classify the SWS database would have
essentially  ignored the additional information gained from the 
larger bandwidth, higher spectral resolution, and greater sensitivity of SWS.

\subsection{Clarifying Evolutionary Patterns}

\subsubsection{The CO Paradigm \label{sec.co}}

The search for patterns and relations among the infrared spectral 
classifications identified here must first consider the observed
dichotomy between carbon-rich and oxygen-rich dust chemistry.
In evolved stars, the chemistry of the dust depends on the C/O
ratio of the material ejected from the envelope.  The formation of
CO molecules will exhaust the less abundant of carbon or oxygen,
leaving the other element available to form molecules which serve as
seeds for dust formation.  This CO paradigm works
admirably well.  In only a few cases does the chemistry of the stellar 
photosphere differ from that of the dust, and most of these cases probably
arise within binary systems.  For example, the silicate carbon stars (2.C/SE)
show carbon-rich photospheric features and oxygen-rich dust.  In
these sources, the dust emission may arise from a disk around an
unseen companion which trapped mass lost from the primary before it
evolved into a carbon star \citep[see][for a recent study of the SWS
spectrum of V778 Cyg and a discussion of competing
models]{yam00}.

Circumstellar dust shells form in relatively pure environments, but 
interstellar dust represents a mixture of material ejected by many generations 
of evolved stars with a wide variety of progenitor masses.  Since oxygen-rich 
dust shells outnumber carbon-rich dust shells, oxygen-rich dust dominates in 
the interstellar medium.  This means that an oxygen-rich dust spectrum can 
arise in either a pre-main-sequence or a post-main-sequence environment, but 
carbon-rich spectra will only appear in post-main-sequence objects.

\subsubsection{Oxygen-Rich Dust Emission \label{sec.orich}}

The oxygen-rich post-main-sequence objects can be organized along the sequence
from AGB source $\rightarrow$ OH/IR source $\rightarrow$ PPN
$\rightarrow$ PN.  This sequence assumes that as the average
oxygen-rich star evolves up the AGB, its mass-loss rate increases,
eventually enshrouding it so deeply within its circumstellar
dust shell that it disappears completely from the optical sky.  This
first stage of development is well-documented.  \cite{jon90}, in a
study of variable AGB sources identified by the Air Force Geophysics
Laboratory (AFGL) infrared sky
survey \citep{pm83}, showed that as Miras evolve to OH/IR stars, the
period of variability and mass-loss rate steadily increase as the
infrared colors progressively redden.  They showed photometrically
that this evolution transformed the silicate emission feature at
10~\mum\ to a deep absorption feature.  \cite{le90} illustrates
this change spectroscopically with LRS data in his Figure 3.

Examining the composition of each subgroup (in terms of the fraction
represented by AGB sources, OH/IR stars, PPNe, and PNe) helps to
organize the spectral subgroups defined in our classification
system into an evolutionary sequence.  In some cases, the
composition of a subgroup is obvious; in others the small sample
sizes and the inherent selection effects of the SWS database limit
the usefulness of this method.

The initial steps are relatively straightforward to interpret.
A star on the early AGB will appear as a naked star with absorption
bands from CO and SiO (1.NO).  Reinterpreting Figure 8 of \cite{sp95}
in terms of the subgroups defined here, the shift from 1.NO to 2.SE
occurs between (time-averaged) spectral types of M4 and M5.  This
marks the onset of significant mass loss and dust formation, but the
detailed evolution through the various classes of silicate emission
(broad, structured, and classic) is more difficult to trace.
\cite{sp95,sp98} found few correlations between spectral
type, variability class, and the shape of the silicate dust
spectrum.  They suggested that the formation of multiple shells
might cloud the picture, and that the shape of the spectrum might
depend on photospheric C/O ratio, which would imply that dredge-ups
of processed material from the stellar interior might determine the
shape of the infrared dust features. Detailed analysis of the shape
of the silicate feature and related features, such as the CO$_2$ lines and
the 13 and 19.5 \mum\ bumps, in the 2.SE subgroups may shed further
light on this subject \citep{sgkp02}.

As the dust contribution grows to dominate the stellar spectrum, the
spectrum will shift from Group 2 to Group 3 (3.SE).  It will then
develop into a 3.SB spectrum as the optical depth of the silicate
dust increases and drives the 10~\mum\ feature into self-absorption.
The 3.SE sources are a mixture of M stars on the AGB, M supergiants,
and optically enshrouded OH/IR stars.  The 3.SB sources are more
evolved, with later spectral types and longer periods of
variability.  All of the sources in 3.SE were initially discovered in
infrared surveys, and all show OH masers.  The lack of a single
source with a 2.SB spectrum suggests that the
transition to self-absorption occurs in Group 3.

Evolution continues from 3.SB to 4.SA, where the silicate feature
goes into full absorption and the SED becomes redder.  The sources
in this group are associated with OH/IR stars, many of them Mira
variables, and PPNe.  The transition to SA does not occur within
Group 3, because all of the 3.SA spectra are associated with
pre-main-sequence sources.  It also does not appear to occur
frequently within Group 4, as most of the 4.SB sources appear to
be much more evolved PNe.  Rather, the transition from SB to SA
appears to coincide with the transition from Group 3 to Group 4.

The stages following 4.SA are much less clear.  Ultimately, the
high mass-loss rates associated with the end of the AGB-OH/IR phase
will strip the envelope from the core, producing a PPN.  As the
remnant shell expands and thins, revealing the ionized central regions, 
 the source becomes a PN.
How does this process manifest itself into the subgroups not yet
included in the sequence:  4.SB, 4.SE, 4.SEC, and 4.SC?
All four subgroups show roughly the same percentage of clearly
identified PNe (58-60\% of the sample, excluding high-mass
objects and pre-main-sequence objects), but only 4.SC and 4.SEC
include any sources identified as OH/IR stars or still on the AGB.
Because of this,  we suspect that 4.SC and 4.SEC
precede 4.SE and 4.SB on a typical evolutionary path.

\cite{wat96} first identified crystalline species of silicates in
the spectra of circumstellar dust shells associated with evolved
stars using data from the SWS.  They noted that the crystalline
features do not appear until the color temperature of the shell
has decreased to $\sim$300 K.  Further study of SWS data by
\cite{syl99} relates the presence of crystalline features with
the optical depth at 10~\mum.  They suggest that crystalline
silicates do not appear until the mass-loss rate has crossed a
certain threshold.  Thus as mass-loss rate increases, absorption
strength at 10~\mum\ grows stronger, and color temperature reddens,
a typical source will evolve to SA and then to SC.

Most of the sources in subgroups 4.SC, 4.SEC, 4.SB, and 4.SE appear
in the upper middle of the HR diagram (spectral class B, A, F, and G,
usually with emission lines, luminosity class I-II).  Whatever their
precise order, most or all of the post-main-sequence sources
with these classes of spectra are obviously in transition from the
AGB or red supergiant phase to later stages of evolution.  It is 
likely that the difficulty in ordering these subgroups results from
the wide range of stellar masses which can produce oxygen-rich dust
shells (from less than 1 M$_{\sun}$ to beyond 50 M$_{\sun}$).  The more
massive stars do not follow the standard evolutionary scenario;
instead they evolve onto the super-AGB \citep[e.g.][]{gbi94}.  Initial
masses $\ge$
11 M$_{\sun}$ produce final core masses beyond the Chandrasekhar limit
and become supernovae.  Masses $\ga$ 50 M$_{\sun}$ are associated with
the luminous blue variables \citep[e.g.][]{hd94}, some of which are in the SWS 
sample.  With all of these sources producing oxygen-rich dust shells,
perhaps it is not a surprise that the redder spectra cannot be 
ordered into a smooth sequence. 

Another complication is the mixture of young and old sources in
Groups 3 and 4 (in contrast to the oxygen-rich spectra in Groups 1
and 2 (1.NO and 2.SE), most of which are evolved sources).  In
subgroup 4.SE, 9 of 24 sources are clearly pre-main-sequence; all are
Herbig Ae/Be stars except for one source classified as F0e.  This
represents the majority of the young sources in the sample, but three
more Herbig Ae/Be stars appear in subgroup 3.SE (out of 21), both 3.SA
spectra are pre-main-sequence Be stars, subgroups 4.SEC and 4.SB
each contain two pre-main-sequence sources (out of 10 and 7
respectively), and one of the 14 4.SC sources is young (a T Tauri
star).  Three of the four young sources in subgroups 4.SEC and 4.SB
are Herbig Ae/Be stars; the other source is an Ae star.

It is unfortunate that young and old sources appearing in the
same part of the HR diagram, luminous Be, Ae, Fe, and G stars,
exhibit similar infrared spectral characteristics.  Determining
whether these sources were evolving {\em to} or {\em from} the
main sequence has been a long-standing problem in astronomy. 
\cite{wcv89} showed that some types of young and old stars could be 
separated into different zones using \iras\ color-color diagrams. 
As Figures 4 and 5 show, the SWS spectra extend sufficiently beyond 20 \mum\
for the shape of the continuum to be defined, thus showing the underlying
dust temperature. The shape of the dust continuum from $\sim20$ to $\sim40$
\mum\ might be one way of separating the young and old objects.
Potentially, the more detailed Level 3 classification will address this issue.

\subsubsection{The Carbon-Rich Dust Sequence \label{sec.crich}}

While oxygen-rich dust can occur in both evolved stars and
environments associated with star formation, carbon-rich dust only
occurs in the vicinity of carbon stars or in planetary nebulae which
have presumably evolved from carbon stars.  Furthermore, the range of
stellar masses which evolve to carbon stars is limited to $\ga$
2 M$_{\sun}$ and less than several M$_{\sun}$ \citep[cf.][and 
references therein]{wk98}.  Perhaps for
these reasons, the carbon-rich spectral classes defined here fall
into a reasonably ordered evolutionary sequence:  1.NC $\rightarrow$
2.CE $\rightarrow$ 3.CE $\rightarrow$ 3.CR $\rightarrow$ 4.CR
$\rightarrow$ 4.CN.

As the mass-loss rate from a naked star with a carbon-rich
photosphere (1.NC) grows, its infrared spectrum develops a strong
emission feature at $\sim$11.5~\mum\ from SiC, producing a 2.CE
spectrum.  Further increases in mass-loss rate lead to a cooler,
optically thicker shell which enshrouds the central star.  The
spectrum is then classified as 3.CE.  It next evolves to 3.CR as
the emitting layer of the dust shell cools and amorphous carbon
begins to dominate the spectrum.  Further thickening of the dust
shell shifts the spectrum to 4.CR.  The next stage is less certain
because the relation of 4.CT to the sequence is not clear.  Perhaps
spectra evolve from 4.CR to 4.CN (i.e. to carbon-rich PPNe), and
only some unusual circumstances lead to the development of a 4.CT
spectrum.  Possibly,  all carbon-rich sources pass briefly through
this stage.  However, the latter possibility seems unlikely given the 
difficulty of fitting the 4.CT spectra into the rest of the carbon
sequence.

\section{Summary}

We examined and categorized the entire \iso-SWS database of 1248 SWS01 
full-grating spectra. A comprehensive spectral classification system 
was developed according to the shape of the SED, that is, the temperature 
of the strongest emitter. Groups were further subdivided based on spectral
features such as silicate emission, ice absorption, or fine-structure lines. 
Most sources which had LRS-based classifications
are in similar categories based on their SWS spectra. Where discrepancies 
occur, e.g, in carbon- vs. oxygen-rich or red vs. blue SEDs, the SWS 
classification should take 
precedence because of the larger bandpass, higher resolution (spectral and 
angular), and greater sensitivity of  SWS.  
As the Level 3 effort progresses, some shifting of individual sources may 
occur, but the overall classification system should be robust.

\acknowledgments

KEK would like to thank the National Research Council for support via a 
Research Associateship through the Air Force Office of Scientific Research. 
This work was supported in part by a NASA grant for the analysis of \iso\ 
dedicated time observations. SDP acknowledges the efforts of Thijs
de Graauw in conducting this experiment.  As the SWS Principal Investigator, 
his advocacy for
the STARTYPE and follow-on related experiments were critical in obtaining the
spectra of sources in the ``missing'' classes.  SDP also thanks Thijs
de Graauw, Harm Habing and Martin Kessler for contributing a portion of their
allocated observing time to this experiment.  Timo Prusti helped
considerably by providing the dedicated and open time SWS01 observing
lists that allowed us to efficiently retarget the STARTYPE sources.  Russ
Shipman provided helpful insight into the calibration of the SWS Interactive
Analysis software and provided an early version of OLP 10.0.
This research has made use of NASA's Astrophysics Data
System Abstract Service, SIMBAD, and the on-line Dictionary of Nomenclature of 
Celestial Objects of the CDS.

\appendix

\section{The Classifications}

Table \ref{tab.data} contains the Group and Subgroup classification
for each source. Sources are ordered by increasing right ascension.
 Columns 1 and 2 contain the source identification
and TDT number of the observation; the observed RA and Dec (J2000)
are given in columns 3 and 4.  The classification is in column 5, and
comments found in the table notes are in column 6.
The coordinates are as given by the observer and represent the
nominal telescope pointing.  They can differ by up to
several arcseconds with respect to the nominal coordinates of
a given object.

The comments can contain important information regarding the 
reliability of the spectra and their classifications. The
most important two comments are ``F'' and ``W.'' ``F'' indicates that
a quality flag was attached to the data, either telemetry, pointing,
or unknown.  Of the 34 flagged observations, 19 could be classified
whereas the rest are in Group 7, and are probably irrecoverable.
Sources with ``W'' in the comment column were observed at the wrong
coordinates.  Two of these were classified based on the detected
flux, although a well centered observation might produce a different
classification.  Note that
an observation does not get a ``W'' if the observer simply mislabelled
the object.

\subsection{Source Names}

Given the heterogeneity of the source names in the IDA, not to 
mention the inaccuracy or non-standard nature of some names
(e.g. GL989 [sic] for AFGL 899), the source identification is not
necessarily that given by the original observer.  Coordinates for each
observation were submitted to SIMBAD for a list of all objects
within 30\arcsec.  Although this process is subject to the errors 
 known to be in SIMBAD, it succeeded in identifying most 
``Off'' or ``Reference'' positions, as well as those sources with
incorrect coordinates (e.g., the observer submitted B1950
instead of J2000 coordinates).

Generally, we preferred older catalog names  over newer designations.
For example, we used the Greek+constellation designation
over a variable star name,  HR over HD, and HD over SAO.  Similarly,
for nebulae, the Messier number takes precedence over NGC or IC
names.  However, there are exceptions. If a newer name contains useful 
information,
it might be used instead of an older one.  For instance, WR (Wolf-Rayet
stars) and MWC (emission line stars) numbers are used instead of HD
numbers as appropriate.  Another exception to the age preference are
Flamsteed numbers, which were generally avoided unless the source is
commonly referenced in the literature by that name (5 sources).  When
in doubt as to what to call a source, we tried to follow the most common
usage in the literature as a guide.

A number of observations have no apparent counterpart in the SIMBAD 
database.  In these cases (37), the names given by the observer are
used, and their origin is indicated by a notation (``Propn'') in the Comments
column of the Table.  Objects in the Galactic center have ``GC''
prepended to their designations.  An example of both these situations
is GC SE\_NTF\_Xng, where the observer called this position
SE\_NTF\_Xng, presumably a non-thermal filament crossing in the
southeast; adding the ``GC'' indicates it is a Galactic center
object. 

There are also observations of the same source at different
positions.  These objects were typically either calibration sources
such as $\gamma$ Dra and $\alpha$ Boo, or extended objects such as
Cas A or M 17. A nominal (0,0) position was chosen for each source.
The offsets for a particular observation are then included in the
name. For example, $\alpha$ Boo $-0.39$, +3.4 is
$(\Delta\alpha,\Delta\delta)=(-0\fs39, +3\farcs4)$ from $\alpha$ Boo.
Nineteen objects and 91 observations have this type of name, and
are noted with ``Offset'' in the comment column.

\begin{deluxetable}{lcccccccccc}
\tablecaption{\iras\ Population Coverage \label{tab.irascnts}}  
\tablewidth{0pt}
\tabletypesize{\small}
\rotate
 \tablehead{
& \multicolumn{10}{c}{Subclass} \\
\colhead{Class}&\colhead{0}&\colhead{1}&\colhead{2}&\colhead{3}&\colhead{4}&\colhead{5}&\colhead{6}&\colhead{7}&\colhead{8}&\colhead{9}\\
}
\startdata
\cutinhead{LRS Atlas Classifications}
  0  &    0,1/1 &  6,21/313&  3,3/4  &  0,1/1  &  1,5/12 &  8,6/32 & \\
  1  &    1,1/2 &   0,1/1 &   2,4/20 & 3,15/116& 7,34/324& 19,22/480 & 22,20/390& 11,16/349& 40,21/460&  7,4/96 \\
  2  &        &  9,10/43&  16,23/151&  9,13/137& 10,16/155& 5,14/163 & 9,17/175& 8,14/197& 8,14/210& 12,27/499\\
  3  &        &  9,10/46&  5,11/49 &  1,4/30 &  0,6/22 &  2,4/19 &  0,3/9  &  1,1/8  &  4,3/8  &  6,4/39\\
  4  &        &   6,6/26&  15,16/152& 6,13/133& 9,10/113&  5,6/64 &  1,2/21 &  1,2/15 &  0,1/3  &  1,2/11\\
  5  &   6,11/51&   2,3/4 &   0,2/4  &  1,4/4  &       &	&      &	      &	      &\\
  6  &        &       &   2,1/2  & 0,0/3  &  1,1/6  &  0,0/3  &  1,1/4  &  1,1/7  &  0,0/3  & 9,16/50\\
  7  &        &   1,4/7 &   4,4/9  &  3,2/9  &  3,2/7  &  2,1/8  &  3,2/5  &  1,0/3  &  0,0/1  &  4,4/18\\
\enddata
\tablecomments{In each class, the numbers given $a,b/c$ are 
$a$ the number of sources actually observed, $b$ the number selected for 
observation by STARTYPE, and $c$ the total number in the subclass.}
\end{deluxetable}

\begin{deluxetable}{ll}
\tablecaption{Level 2 Classification Definitions \label{tab.abbrev}}
\tablewidth{0pt}
 \tablehead{
\colhead{Class}&\colhead{Description}\\
}
\startdata
SE   & {\bf S}ilicate (or oxygen-rich) dust {\bf E}mission
       (10--12 and 18--20 \mum) \\
SB   & {\bf S}ilicate emission in self-a{\bf B}sorption 
       (10 \mum) \\
SA   & {\bf S}ilicate {\bf A}bsorption (10--12 \mum) \\
SC   & {\bf S}ilicate emission from {\bf C}rystalline grains 
       (33, 40, 43 \mum) \\
SEC  & {\bf S}ilicate {\bf E}mission from {\bf C}rystalline grains 
       (11, 19, 23, 33 \mum) \\
CE   & {\bf C}arbon-rich dust {\bf E}mission,
       primarily from SiC (11.5 \mum) \\
CR   & {\bf C}arbon-rich dust emission in a {\bf R}eddened shell\\
     &	 (with features at 11.5 and 26 \mum, often 13.7 \mum\ absorption)\\
CT   &  8, 11.5, {\bf 21}, 26 \mum, no 13.7 absorption\\
CN   & {\bf C}arbon-rich {\bf N}ebulae \\
C/SE & {\bf C}arbon-rich, plus {\bf S}ilicate emission (10--12 \mum)\\
C/SC & {\bf C}arbon-rich, plus {\bf c}rystalline {\bf s}ilicate emission  \\
U/SC & {\bf C}rystalline {\bf s}ilicate and {\bf U}IR emission features \\
U    & Prominent {\bf U}IR emission features\\
PN   & Many prominent atomic fine-structure lines typical of 
{\bf P}lanetary {\bf N}ebulae\\
PU   & As PN, but with strong {\bf U}IR emission \\ 
W    & Emission peaks 6-8 \mum\ \\
F    & Basically {\bf F}eatureless \\
E    & Strong {\bf E}mission lines \\
M    & {\bf M}iscellaneous\\
\enddata
\end{deluxetable}

\begin{deluxetable}{ll}
\tablecaption{Level 2 Suffixes \label{tab.suf}}
\tablewidth{0pt}
 \tablehead{
\colhead{Suffix}&\colhead{Description} \\
}
\startdata
e & Emission lines (e.g. H-recombination, atomic fine-structure, etc.) \\
u & UIR features present, but not dominant feature  \\
p & Fits in given category, but is peculiar \\
:, :: & Uncertain (either noisy or odd) \\
\enddata
\end{deluxetable}

\begin{deluxetable}{lccccccc}
\tablecaption{Level 1-2 Cross-References and Statistics\label{tab.cross1-2b}}
\tablewidth{0pt}
 \tablehead{
& \multicolumn{5}{c}{LEVEL 1} \\
\colhead{LEVEL 2}  & \colhead{Group 1} & \colhead{Group 2} & \colhead{Group 3} & \colhead{Group 4} & \colhead{Group 5}& \colhead{Group 6}& \colhead{Group 7} \\
}
\startdata
N    & 49/59 \\
NO   & 49/90 \\
NC   & 4/7 \\
SE   &    &a: 59/72\tablenotemark{a}  & 21/27    & 25/27    & 7/7  \\
     &    &b: 26/43 \\
     &    &c: 42/53 \\
SB   &    &\nodata   & 9/12     & 7/8      &\nodata    \\
SA   &    &\nodata   & 2/3      & 25/30    & 50/63  \\
\enddata
\tablecomments{The format of the entries is Sources/Spectra.
``Spectra'' refers to the total number of observations in a
particular category. ``Sources'' refers to the number of 
spatially distinct objects, although this can be debatable in cases of 
extended, complex sources. For example, the 16 observations of Cas A 
are counted as one source. Group 7 Sources include only those that
are not offs (31), flagged (13), or at the wrong coordinates (28).}

\tablenotetext{a}{Subroup 2.SE is divided into parts a, b, and c (see text). }

\end{deluxetable}

\begin{deluxetable}{llll}
\tablecaption{KSPW vs. LRS Classifications\label{tab.irascomp}}  
\tablewidth{0pt}
\tabletypesize{\footnotesize}
\rotate
 \tablehead{
 \colhead{LRS}  & \multicolumn{3}{c}{KSPW} \\
                & \colhead{OK} & \colhead{Information Lost}  & \colhead{Mismatch} \\
}
\startdata
``Blue'' \\
$1n$ Featureless & 1.N & 1.NC, 1.NO (1.NE) & 2.CE, 2.M, 2.SEa, 2.SEb, 3.CR, 3.SA, 3.W, 4.CR, 4.SB, 4.U/SC, 6 \\
$2n$ 10 \mum\ Emission & 2.SEa, 2.SEb, 2.SEc, 3.SB, 3.SE & 2.C/SE & 1.N, 2.CE,  4.CR, 4.SE, 4.SEC \\
$3n$ 10 \mum\ Absorption & 2.M, 3.SB, 3.W (3.SA) & & 1.NC, 2.SEb, 4.SA, 4.SB, 4.SC, 5.SA, 5.U, 6 \\
$4n$ Carbon-rich & 2.CE & 3.CE, 3.CR,  4.CR & 2.SEa, 3.SB, 4.SB\\
                 && (4.CN, 4.CT, 4.C/SC) \\
``Red'' \\
$5n$ Featureless & 4.F (5.F) & (4.SC) & 2.SEa, 3.CR, 4.C/SC, 4.SB, 5.SA, 5.UE \\
$6n$ 10 \mum\ Em. & 3.SE, 4.SE, 5.SE (4.SB) & 4.SEC & 2.SEc, 4.CN, 4.SC \\
$7n$ 10 \mum\ Abs. & 4.SA, 5.SA (3.SA, 4.SB) & 4.U/SC & 4.F, 4.CN, 4.CT, 4.SC, 5.U, 5.UE\\
$8n$ UIR + lines & 4.M, 4.PU, 5.U, 5.UE &  2.U, 4.PN \\
$9n$ lines only  & 4.PN, 5.E, 5.PN & 4.PU, 4.SEC, 5.F, 5.SA, 5.UE\\
\enddata
\tablecomments{KSPW classes in parentheses could reasonably have appeared in a given LRS class but did not.
}
\end{deluxetable}

%\documentclass[12pt, preprint]{aastex}
%\begin{document}

\begin{deluxetable}{rrrrlc}
\tablecaption{Source Classification  \label{tab.data}}
\tablewidth{0pt}
\tabletypesize{\small}
 \tablehead{
\colhead{Name} & \colhead{TDT}& \colhead{RA (J2000)} 
& \colhead{Dec (J2000)} &\colhead{Group}& \colhead{Comments}\\
}
\startdata
W Cet            & 37802225       & $00^h02^m07\fs70$        & $-14\arcdeg40\arcmin35\farcs9$  & 2.SEa: &          \\
SV And           & 42801007       & 00 04 20.00         & +40 06 37.2             & 2.SEa:  &          \\
SV And           & 80800708       & 00 04 20.00         & +40 06 37.2             & 2.SEa   &          \\
CIT 1            & 78201008       & 00 06 52.30         & +43 04 36.0             & 7       &  W       \\
HR 10            & 37802001       & 00 07 18.20         & $-$17 23 13.2           & 1.NM:   &          \\
$\beta$ Cas      & 28501420       & 00 09 10.47         & +59 08 59.8             & 1.N     &          \\
V633 Cas         & 43501514       & 00 11 26.60         & +58 50 04.0             & 5.SE    &  W       \\
NGC 40           & 44401917       & 00 13 00.91         & +72 31 20.0             & 4.PN    &          \\
\enddata
\tablecomments{Comments include: F = quality flag such as pointing or telemetry
problems; G = extragalactic source; W = wrong coordinates; o: probably an off, but odd in some way;
R, R: = possibly a recoverable group 7; H = probably irrecoverable in 
group 7; Offset = name includes offsets from nominal position indicated by (0,0); Propn = Name from original observer}
\tablenotetext{a}{ M 1: (0,0) position is $\alpha,\delta=05~34~31.97,~ +22~ 00~ 52.1$ (J2000; Han \&  Tian 1999);
M 17: (0,0) position is $\alpha,\delta=18~20~24.83,~-16~11~34.9$ (J2000, Johnson et al. 1998);
Cas A: (0,0) position  is the Chandra point source $\alpha,\delta=23~23~27.94,+58~48~42.4$ 
(J2000; Tananbaum 1999).}
\end{deluxetable}

%\end{document}
%\include{table2bshell}

\end{document}